\newif\ifpaper
\newif\ifref
\papertrue         
\reffalse          

\ifpaper
  \ifref
    \documentstyle[graphics,referee,endfloat,astrobib,amsmath]{mn-ab}
    
  \else
    \documentstyle[graphics,astrobib,amsmath,fixfloats]{mn-ab}
  \fi

  \ifCUPmtlplainloaded \else
    \DeclareSymbolFont{UPM}{U}{eur}{m}{n}
    \SetSymbolFont{UPM}{bold}{U}{eur}{b}{n}
    \DeclareSymbolFont{AMSa}{U}{msa}{m}{n}
    \DeclareMathSymbol{\upi}{0}{UPM}{"19}
    \DeclareMathSymbol{\umu}{0}{UPM}{"16}
    \DeclareMathSymbol{\upartial}{0}{UPM}{"40}
    \DeclareMathSymbol{\leqslant}{3}{AMSa}{"36}
    \DeclareMathSymbol{\geqslant}{3}{AMSa}{"3E}
  \fi

  \title[CDM-variant models -- I]
        {CDM-variant cosmological models -- I: Simulations and preliminary
         comparisons}
  \author[M. A. K. Gross et al.]
         {Michael~A.~K.~Gross,$^{1,2}$\footnotemark[1]
          Rachel~S.~Somerville,$^{1,3}$\footnotemark[1]
          Joel~R.~Primack,$^1$\footnotemark[1]
          \newauthor
          Jon~Holtzman$^4$\footnotemark[1] and
          Anatoly~Klypin$^4$\footnotemark[1]\\
          $^1$Physics Department, University of California, Santa Cruz, CA 95064 USA\\
          $^2$Earth \& Space Data Computing Division, Code 931, NASA/Goddard Space Flight Center, Greenbelt, MD 20771 USA\\
          $^3$Racah Institute of Physics, The Hebrew University, Jerusalem 91904, Israel\\
          $^4$Department of Astronomy, New Mexico State University, Las Cruces, NM 88001 USA\\
         }

\newcommand{\bvec}[1]{\mbox{\boldmath$\vec{#1}$}}

\newcommand{\Nvir}{N_{\mathrm{vir}}}
\newcommand{\Mvir}{M_{\mathrm{vir}}}

\newcommand{\rvir}{r_{\mathrm{vir}}}

\newcommand{\dvir}{\delta_{\mathrm{vir}}}

\newcommand{\omk}{\Omega_{\mathrm{k}}}
\newcommand{\oml}{\Omega_{\mathrm{\Lambda}}}
\newcommand{\omc}{\Omega_{\mathrm{c}}}
\newcommand{\omh}{\Omega_{\mathrm{\nu}}}
\newcommand{\omb}{\Omega_{\mathrm{b}}}

\newcommand{\rhocrit}{\rho_{\mathrm{c}}}
\newcommand{\rhoh}{\rho_{\mathrm{h}}}

\newcommand{\delc}{\delta_{\mathrm{c}}}

\newcommand{\Np}{N_{\mathrm{p}}}
\newcommand{\Ng}{N_{\mathrm{g}}}

\newcommand{\delcg} {\delta_{\mathrm{c,g}}}
\newcommand{\delct} {\delta_{\mathrm{c,t}}}
\newcommand{\rmin} {r_{\mathrm{min}}}
\newcommand{\mmin} {M_{\mathrm{min}}}
\newcommand{\nmin} {N_{\mathrm{min}}}

\newcommand{\hkpc}{\mbox{$h^{-1}$ kpc}}
\newcommand{\hmpc}{\mbox{$h^{-1}$ Mpc}}
\newcommand{\hmsun}{\mbox{$h^{-1}$ $M_{\odot}$}}
\newcommand{\hmpcinv}{\mbox{$h$ Mpc$^{-1}$}}
\newcommand{\kms}{\mbox{km s$^{-1}$}}
\newcommand{\kmsmpc}{\mbox{km s$^{-1}$ Mpc$^{-1}$}}
\newcommand{\potent}{\textsc{potent}\ }

\newcommand{\adot}{a^{-1/2}\sqrt{\omc+\omh+\oml a^3+\omk a}}

\newcommand{\cobe} {\emph{COBE}}

\def\la{\mathrel{\hbox{\rlap{\hbox{\lower4pt\hbox{$\sim$}}}\hbox{$<$}}}}
\def\ga{\mathrel{\hbox{\rlap{\hbox{\lower4pt\hbox{$\sim$}}}\hbox{$>$}}}}

\renewcommand{\topfraction}{1}
\renewcommand{\dbltopfraction}{1}
\renewcommand{\bottomfraction}{1}
\renewcommand{\textfraction}{0}

\newenvironment{fig}[1]{%
    \begin{center}
      \ifpaper
        {\Large\resizebox*{\linewidth}{!}{{\input #1 }}}
      \else
        {\Large\resizebox*{\textwidth}{!}{{\input #1 }}}
      \fi
    \end{center}
  }
  {
}
\newenvironment{figwide}[1]{%
    \begin{center}
      \resizebox*{\textwidth}{!}{{\input #1 }}
    \end{center}
  }
  {
}
\newenvironment{figlong}[1]{%
    \begin{center}
      \resizebox*{!}{0.75\textheight}{{\input #1 }}
    \end{center}
  }
  {
}

\newsavebox{\tblbox}
\newlength{\tblwidth}
\newcommand{\tblcapt}[4]{%
    \sbox{\tblbox}{#1}
    \settowidth{\tblwidth}{\usebox{\tblbox}}
    \hspace*{\fill}
    \begin{minipage}{\tblwidth}
      \caption{#2}
      \label{#3}
      \usebox{\tblbox}\\[0.5ex]
      {\def\baselinestretch{1.0} \Huge\small #4}
    \end{minipage}
    \hspace*{\fill}
}

\newcommand{\tblcaptraw}[4]{%
    \sbox{\tblbox}{#1}
    \settowidth{\tblwidth}{\usebox{\tblbox}}
    \begin{minipage}{\linewidth}
      \caption{#2}
      \label{#3}
      \hspace*{\fill}
      \begin{minipage}{\tblwidth}
        \usebox{\tblbox}
      \end{minipage}
      \hspace*{\fill}\\[0.5ex]
      {\def\baselinestretch{1.0} \Huge\small #4}
    \end{minipage}
}
\newcommand{\tblcaptland}[4]{%
    \ifpaper \else
       \rotatebox{90}{%
    \fi
    \sbox{\tblbox}{#1}
    \settowidth{\tblwidth}{\usebox{\tblbox}}
    \hspace*{\fill}
    \begin{minipage}{\tblwidth}
      \caption{#2}
      \label{#3}
      \usebox{\tblbox}\\[0.5ex]
      {\def\baselinestretch{1.0} \Huge\small #4}
    \end{minipage}
    \hspace*{\fill}
    \ifpaper \else
       }
    \fi
}

\begin{document}
  \maketitle
\else
  \chapter{Nonlinear comparisons of CDM-variant cosmological models}
\fi

 
\ifpaper \, \else
    \chapter[Simulations and preliminary comparisons]
            {Simulations and preliminary comparisons}
    \label{chap:suites}
\fi


\ifpaper
\begin{abstract}
\else
\chapterabstract
\fi
We present two matched sets of five dissipationless simulations each, 
including four presently favored minimal modifications to the standard cold
dark matter (CDM) scenario.  One simulation suite, with a linear box size of 75
$\hmpc$, is designed for high resolution and good statistics on the group/poor
cluster scale, and the other, with a box size of 300 $\hmpc$, is designed for
good rich cluster statistics.  All runs had 57 million cold particles, and
models with massive neutrinos (CHDM-2$\nu$) had an additional 113 million
hot particles.  We consider separately models with massive neutrinos, tilt,
curvature, and a nonzero cosmological constant ($\Lambda\equiv 3H_0^2\oml$) in
addition to the standard CDM model.  We find that the dark matter in each of
our tilted $\Omega_0+\oml=1$ (T$\Lambda$CDM) model with $\Omega_0=0.4$, our
tilted $\Omega_0=1$ model (TCDM), and our open $\Lambda=0$ (OCDM) model with
$\Omega_0=0.5$ has too much small-scale power by a factor of $\sim 2$, while
CHDM-2$\nu$ and SCDM are acceptable fits.  In addition, we take advantage of
the large dynamic range in detectable halo masses afforded by the combination
of the two sets of simulations to test the Press-Schechter approximation.  
We find good fits at cluster masses for $\delcg=1.27$--$1.35$ for a Gaussian
filter and $\delct=1.57$--$1.73$ for a tophat filter. But, when we adjust
$\delc$ to obtain a good fit at cluster mass scales, we find that the
Press-Schechter model overpredicts the number density of halos compared to the
simulations by a weakly cosmology-dependent factor of 1.5--2 at galaxy and
group masses. It is impossible to obtain a good fit over the entire range of
masses simulated by adjusting $\delc$ within reasonable bounds.
\ifpaper
  \end{abstract}

  \begin{keywords}
    large-scale structure of universe -- dark matter --
    cosmology:theory -- cosmic microwave background
  \end{keywords}
  \renewcommand{\thefootnote}{{\fnsymbol{footnote}}}
  \footnotetext[1]{E-mail:~gross@fozzie.gsfc.nasa.gov~(MAKG);%
                           \ifref \else \\ \fi
                           rachels@alf.fiz.huji.ac.il~(RSS);
                           joel@ucolick.org~(JRP);\\
                           holtz@nmsu.edu~(JH);
                           aklypin@nmsu.edu~(AK)}
  \renewcommand{\thefootnote}{{\arabic{footnote}}}
\fi


\section{Introduction}

The \cobe\ DMR detection of anisotropies in the cosmic microwave
background \cite{smoot:1992} made it very clear that the `standard' structure
formation scenario of cold dark matter \cite{blumenthal:1984,DEFW:1985}
cannot simultaneously account for fluctuations on very large and very small
scales.  That model made several
very restrictive assumptions about cosmological parameters -- that spacetime is
homogenous, isotropic and globally flat; that there is no cosmological
constant; that fluctuations from homogeneity are Gaussian-distributed and
nearly scale-independent at horizon crossing; that the Hubble parameter
$h\equiv H_0$ / (100 $\kmsmpc$) is 0.5; and that the number of
free parameters is minimized.  The obvious fixes to the
problem of excess small-scale power (when normalizing power spectra to the
\cobe\ anisotropy) are to make one of the following modifications to the
model:
\begin{enumerate}
\item tilt the primordial spectrum,
\item allow a nonzero cosmological constant but retain globally
flat geometry,
\item allow the universe to be open,
\item add hot dark matter (i.e., neutrinos with masses of a few eV), or
\item lower the Hubble parameter much further ($h\sim 0.3$--$0.4$).
\end{enumerate}
Each of these modifications adds only one free parameter to the cosmology.
In this paper, we consider the most viable models from each class above except
the last, and simulate them with an $N$-body code in two suites, with equivalent
initial conditions across all the models.  We do not consider a
`low-$H_0$' model \cite{bartlett:1995} because of increasingly solid
observational evidence that $h\ga 0.5$.

Deciding on cosmological parameters is to some extent an iterative process.
Much can be done using the Press-Schechter \citeyear{press:1974} approximation,
but the assumptions that go into it are not necessarily realistic (for example,
spherical symmetry -- see \citeNP{jain:1994} and \citeNP{monaco:1995}).
Therefore, it is useful as a \emph{first approximation} to calculating the
mass functions, and we use it to perform an approximate cluster normalization,
using guesses about other cosmological parameters.  We run a set of simulations
and use them to test the Press-Schechter approximation, and make several
preliminary comparisons to observational data.  In
\ifpaper
  a companion paper \cite{gross:clusters},
\else
  chapter~\ref{chap:clusters},
\fi
we recalibrate the Press-Schechter approximation and use it to derive
refined estimates of model normalization and $\Omega_0$ from several different
data sets, and make more careful comparisons to cluster abundance.
Subsequent papers will use simulations based on the refined normalizations.

In section~\ref{sec:models}, we describe our specific models from each class of
CDM-variant models and explain why we chose the parameters as we did.
In section~\ref{sec:algorithm}, we briefly describe the implementation of
the particle-mesh algorithm we used for this study.  We explain our halo
finding algorithm and the effect of mass resolution upon it in
section~\ref{sec:halos} and report the simulation results in
section~\ref{sec:results}.  Finally, in section~\ref{sec:conclusion}, we give
our conclusions.

\section{Simulations}
\label{sec:simulations}

\subsection{Models}
\label{sec:models}

Given the long list of modifications to the cold dark matter scenario in the
previous section, we could construct a model by adjusting \emph{every}
parameter in order to fit all the available  observational data. However, in
addition to being aesthetically displeasing, the physical significance of such
a model would be unclear.  As a result, we have tried to minimize the number of
modifications to the relatively simple standard cold dark matter scenario by
investigating each of the modifications mentioned above in a separate model.
The exceptions to this policy are that in addition to any one of modifications 
\ifpaper
  (ii)--(iv),
\else
  2--4,
\fi
we allow a small tilt, up to $n=0.9$, in order to simultaneously fit the
\cobe\ and cluster data, and we allow the Hubble parameter to be adjusted
within reasonable observational bounds according to the requirements of the
model.  Larger tilts are not allowed because they tend to cause disagreements
with high-multipole cosmic microwave background data.

\begin{figure}
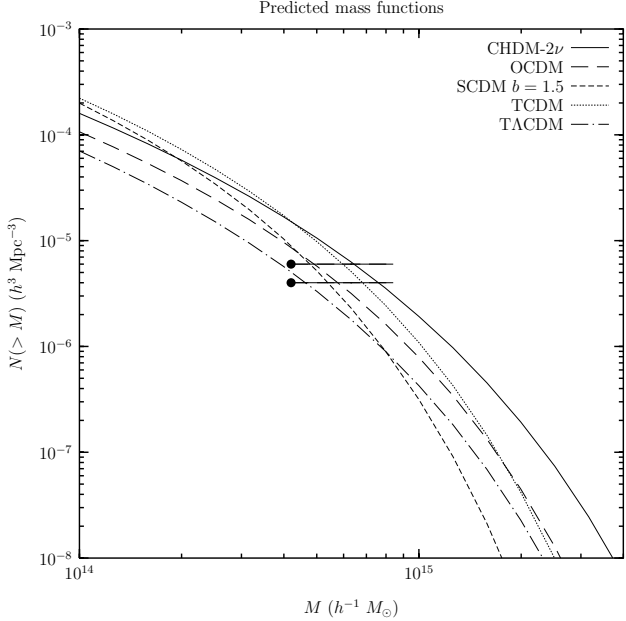

\ifref \else
\begin{fig}{ps}
\fi
  \caption[Expected mass functions for all models]%
          {Expected mass functions for all models estimated from the
           Press-Schechter approximation with a Gaussian filter, using
           $\delcg=1.5$ for the model with massive neutrinos and
           $\delcg=1.3$ for all other models.  The two data points
           shown correspond to observational estimates of cluster abundance
           (WEF93, BGGMM93)\protect\nocite{biviano:1993,white:1993}.  Note that
           cluster density mapping via gravitational lensing
           \protect\cite{squires:1996,squires:1997,miralda-escude:1995,wu:1997}
           may indicate that X-ray masses are systematically low, and the
           masses can plausibly be raised by a factor of up to
           two, which corresponds to the horizontal line on the right of each
           cluster data point.}
  \label{fig:ps}
\ifref \else
\end{fig}
\fi
\end{figure}

We explore the large parameter space by running a large suite of linear
calculations and comparing the output to appropriate observational constraints.
Constraints that we consider in choosing model parameters for more detailed
nonlinear analysis are:
\begin{enumerate}
\item the abundance of Abell clusters, as measured by X-ray temperature profiles
\cite[hereafter WEF93 and BGGMM93, respectively]{white:1993,biviano:1993}.
We assume that cluster masses may be underestimated by up to a factor of two, 
motivated by results from cluster density mapping with gravitational lensing
(\citeNP{squires:1996,squires:1997,miralda-escude:1995,wu:1996,wu:1997};
figure~\ref{fig:ps}),
\item microwave background anisotropies for $\ell\la 800$
(figure~\ref{fig:cmb}) as measured by several recent CMB detection experiments
(\citeNP{tegmark:1996,netterfield:1997,scott:1996,platt:1997};
figure~\ref{fig:cmb})
\item `bulk flow' peculiar velocity measurements and resulting constraints
on the power spectrum (\citeNP{dekel:1997,kolatt:1997};
figure~\ref{fig:bfpow}).
The linear estimates of these parameters are shown in table~\ref{tbl:linear}
and in figures~\ref{fig:ps}, \ref{fig:cmb}, and~\ref{fig:bfpow} for the
models we consider.
\end{enumerate}

In most of the previous work with modified CDM models, the most `extreme'
values of the model parameters have been chosen (i.e. as far from SCDM as was
considered observationally plausible). For example, low-$\Omega_0$ models
typically have values of $\Omega_0\sim0.2-0.3$. However, in every case, while
solving some of the problems with SCDM, this introduces new problems or
conflicts with other observational constraints. Thus our approach will be
somewhat different. We use our previous experience with linear and non-linear
tests of CDM-variant models, as well as the published results of others, to
find models that represent a `middle ground' between SCDM and the most
extreme version of the particular class of model. In this way, we hope to
choose the `best' rather than the most extreme case, and to identify models
that agree with the widest possible range of observations.

For most models, we presume a baryon abundance of $\omb = 0.025 h^{-2}$,
consistent with the \citeN{tytler:1996} cosmic
deuterium abundance measurement.%
\footnote{Burles \& Tytler (\protect\citeyearNP{burles:1997a,burles:1997b}) have
very recently remeasured the deuterium abundance and found it to be 20 per cent
lower, $0.019\pm0.001$.  This makes a very small change in the power spectrum,
and the most significant effect is to make agreement with high-$\ell$ cosmic
microwave background measurements (figure~\ref{fig:cmb}) more difficult.
The height of the first Doppler peak depends strongly on $\omb$.}
Normalization is accomplished by calculating
low multipoles using an enhanced version of the linear code from
\citeN{holtzman:1989} and comparing to the four-year \cobe\ DMR anisotropy
measurements (\citeNP{gorski:1996a}; G\'orski, private
communication).

For comparison to other studies, we also simulated the standard cold dark
matter (SCDM) model with bias $b=\sigma_8^{-1}=1.5$.  That model is intended to
approximately match observed cluster abundances at the cost of being
inconsistent with the \cobe\ anisotropy measurements. For this model, we
presumed there were no baryons in the Universe, and used the BBKS transfer
function \cite{BBKS:1986} used in previous studies, that is,
\begin{multline}
    P(k)=Ak \frac{[\ln(1+2.34q)]^2}{(2.34q)^2}
\ifpaper\ifref\else\times\\ \fi\fi
    \left(1+3.89q+(16.1q)^2+(5.46q)^3+(6.71q)^4\right)^{-1/2}
\label{eqn:bbks}
\end{multline}
with $q=kh^{-2}$ and $A$ adjusted so that the rms fractional variance in
mass in spheres of radius 8 $\hmpc$ estimated using linear theory
is $\sigma_8=0.667$.

The simplest way to solve the problem of excess power on small scales is by 
`tilting' the spectrum, that is, by changing the $Ak$ factor in equation
(\ref{eqn:bbks}) to $Ak^n$, with $n<1$. 
\begin{table*}
  \tblcaptland{
    \begin{tabular}{l@{\extracolsep{\fill}}cllllllllllrc}
    Model & Age$^a$ & $h^b$ & $\Omega_0$ & $\omc$ & $\omb$ & $\omh$ & $\oml$ & $n^c$ &
    $N_{\mathrm{\nu}}^d$ & $\sigma_8^e$ & $\widetilde\sigma_8^f$ & $V_{50}^g$ & $N_{\mathrm{cl}}^h$\\
    \hline
    observations &    &    &   &     &     &   &   &   & &     &     &375&$5\times 10^{-6}$\\
    1-$\sigma$ errors &&   &   &     &     &   &   &   & &     &     & 85&$2\times 10^{-6}$\\
    \\
    CHDM-2$\nu$  &13.0&0.5 &1.0&0.7  &0.1  &0.2&0.0&1.0&2&0.719&0.719&399&$6\times 10^{-6}$\\
    OCDM         &12.3&0.6 &0.5&0.431&0.069&0. &0.0&1.0&0&0.773&0.581&254&$3\times 10^{-6}$\\
    SCDM         &13.0&0.5 &1.0&1.0  &0.0  &0.0&0.0&1.0&0&0.667&0.667&192&$2\times 10^{-6}$\\
    TCDM         &14.5&0.45&1.0&0.9  &0.1  &0.0&0.0&0.9&0&0.732&0.732&270&$5\times 10^{-6}$\\
    T$\Lambda$CDM&14.5&0.6 &0.4&0.365&0.035&0.0&0.6&0.9&0&0.878&0.572&335&$2\times 10^{-6}$\\
    \hline
    \end{tabular}
  }
  {Model parameters and linear results for both simulation suites.}
  {tbl:linear}
  {$^a$ Time since the Big Bang in Gyr.\\
   $^b$ Presumed Hubble parameter, in units of 100 $\kmsmpc$.\\
   $^c$ `Tilt' of the primordial spectrum; $P(k)\propto k^n$.\\
   $^d$ Number of massive neutrinos presumed.  The equivalent mass of a neutrino
        is $m_{\mathrm{\nu}}=\frac{\omh h^2}{N_{\mathrm{\nu}}}\cdot 92$ eV.\\
   $^e$ rms mass fluctuation in a sphere of radius 8 $\hmpc$.\\
   $^f$ $\widetilde\sigma_8\equiv\sigma_8\Omega_0^{0.46-0.10\Omega_0}$ for
        $\Lambda=0$ models, and $\sigma_8\Omega_0^{0.52-0.13\Omega_0}$ for
        $\oml+\Omega_0=1$ models.  \protect\citeN{eke:1996} calculate
        $\widetilde\sigma_8 = 0.52\pm 0.04$ in order to fit cluster temperatures
        assuming $\beta\equiv\mbox{$\langle KE\rangle_{\mathrm{dm}}$} /
                             \mbox{$\langle KE\rangle_{\mathrm{gas}}$}=1$.
        However, new simulations \protect\cite{frenk:1998} show that
        $\beta\approx1.17$, corresponding to $\widetilde\sigma_8=0.61\pm0.05$.\\
   $^g$ rms velocity in a sphere of radius 50 $\hmpc$ \cite{dekel:1997}.
        Note that the observational value is for one particular 50 $\hmpc$
        sphere around the Local Group, and the simulation values are the rms
        value for a distribution of randomly placed spheres.  These are not
        the same, so we can't use the observations to rigorously define
        confidence limits for the simulation quantities.\\
   $^h$ Estimated number density of clusters of mass $>6\times 10^{14}$
        $\hmsun$, in $h^3$ Mpc$^{-3}$, from Press-Schechter theory with a
        Gaussian window function.  $\delcg$ is 1.5 for CHDM models
        \protect\cite{walter:1996,borgani:1997a} and 1.3 for all other models
        \protect\cite[KPH96]{liddle:1996c}\protect\nocite{klypin:1996a}.
        Masses near the center of the allowable range for cluster data 
        (WEF93, BGGMM93)\protect\nocite{white:1993,biviano:1993} are used.
        Note that uncertainties in the masses of measured clusters mean that
        the masses for which densities were measured could shift coherently
        up to a factor of two above the reported value of $4.2\times 10^{14}$
        $\hmsun$.  This is a naive approach which we use only for our
        first iteration of model parameters. We use more sophisticated
        methods in
     \ifpaper
       \citeN{gross:clusters}.
     \else
       chapter~\ref{chap:clusters}.
     \fi
    }
\end{table*}
However, the price paid is that choosing $n<1$ also reduces the amplitude of
the first `Doppler peak' in the small-angle cosmic microwave background
spectrum. We find that $n\sim0.9$ is the largest allowable tilt that is still
marginally consistent with the large-multipole cosmic microwave background
data. With $n=0.9$, when we \cobe\ normalize the model we find that it tends to
overproduce clusters at $M=6\times 10^{14}$ $\hmsun$ according to the
Press-Schechter estimate, unless we use a rather low value of the Hubble
parameter, $h=0.45$. Although this is not favored by most of the current
observational data, we conclude that this choice of parameters consitutes the
best compromise amongst the observational constraints that we have imposed.

Another fix is to add a little hot dark matter, usually assumed to be in the
form of a massive neutrino. Previously studied versions of this class of model
typically postulate a single species of neutrino with significant mass, and a
fraction $\omh=0.3$ of the critical density in the form of hot dark
matter. This model was ruled out, based on its inability to reproduce the
observed abundances of Damped Lyman $\alpha$ systems (DLAS) at $z\sim3$
\cite{kauffmann:1994,klypin:1994}. Models with lower fractions of hot dark
matter ($\omh=0.2$) are more plausibly consistent with constraints from DLAS
\cite{klypin:1994}, but still have too much small scale power and thus
overproduce clusters at $z=0$.  However, as pointed out by
Primack et al. (\citeyearNP{primack:1995a}; see also
\citeNP{pogosyan:1995a}) if the hot dark matter is divided into two species of
neutrino with equal masses, the power on cluster scales
is reduced by 20 per cent without affecting smaller or larger scales. This
lowers cluster abundances without worsening potential early structure formation
problems (small-scale power) or compromising the
\cobe\ normalization. We find reasonable agreement with observed cluster
abundances with $\omh=0.2$, $N_{\mathrm{\nu}}=2$, $h=0.50$, and $n=1$; or
alternatively, with a small tilt ($n=0.9$) and a higher Hubble parameter
($h=0.6$). We chose the former, based on concerns about the age of the
Universe. However, we ran the simulations before the Hipparcos recalibration of
the age of globular clusters
\cite{reid:1997,gratton:1997,chaboyer:1997}. This constraint has now
been considerably weakened, and $h=0.6$ or even $0.65$ would lead to ages
consistent with the present estimates.

The $\oml\neq0$ class of models has been well studied, typically with
$\Omega_0=0.3$ and $h\sim0.65$--$0.7$. However, analysis of earlier $N$-body
simulations has shown that when non-linear effects are included, this model
produces a power spectrum/correlation function with too high an amplitude on
small spatial scales compared to observations, unless galaxies are strongly
anti-biased with respect to the dark matter (\citeNP{klypin:1996a}, hereafter
KPH96; \citeNP{jenkins:1997}). \citeN{ghigna:1997} have also shown that the
void probability function
for this model is in disagreement with observations. Therefore we have chosen a
model with a slightly higher value of the matter density ($\Omega_0=0.4$) and a
tilt ($n=0.9$) to reduce small-scale power and correlations.

Many observers favor an open cosmology and a high Hubble parameter,
consistent with local density estimates and the Hubble Key Project.
The lowest reasonable value of $\Omega_0$, given initial Gaussian fluctuations
as assumed in all CDM-variant models considered here, is constrained to be
above 0.3 at $\ga\!\! 4\sigma$ confidence (\citeNP{nusser:1993};
cf. also \citeNP{dekel:1994b,bernardeau:1995}).  We adopt $\Omega_0=0.5$ as a
`reasonable' value for OCDM, noting that even this relatively high $\Omega_0$
leads to a power spectrum lower than that indicated by the \potent\ analysis
(\citeNP{kolatt:1997}; see also figure~\ref{fig:bfpow}).  Our linear
code is not capable of determining low multipole cosmic microwave background
fluctuations for OCDM, as it uses a plane wave expansion that is only
appropriate for flat cosmologies.  Instead, we use fitting functions for the
normalization $\delta_H(\Omega_0)$ and the transfer function $T(k)$ given by
\citeN[hereafter LLRV96]{liddle:1996c}.%
\footnote{After we ran this model, LLRV96 was superseded by
\protect\citeN{bunn:1997} and \protect\citeN{hu:1997b}.  Those papers'
$\sigma_8$ values agree to high precision with LLRV96\nocite{liddle:1996c} if
one lowers $\omb$ from 0.025 $h^{-2}$ to 0.015 $h^{-2}$, which
\protect\citeN{bunn:1997} favor anyway.  However, the transfer function
shapes are somewhat different, and the LLRV96\protect\nocite{liddle:1996c}
normalization is to the \cobe\ 2-year data, so the power on scales of a few
hundred $\hkpc$ may be up to 20 per cent low compared to
\protect\citeN{bunn:1997} and \protect\citeN{hu:1997b}.
Using a BBKS-style fit as all three papers do, rather than integrating the
Boltzmann equation directly, introduces an error of similar magnitude, even
with the improved shape parameter described in
\citeN[equation D-29]{hu:1996}.  We therefore neglect the difference
between the \protect\citeN{bunn:1997} and
LLRV96\protect\nocite{liddle:1996c} spectra.}

Figure~\ref{fig:ps}%
summarizes the expected mass functions on the group and cluster mass scales,
as estimated from the Press-Schechter approximation, with a Gaussian filter.  
Using the calibration with $N$-body simulations from \citeN{borgani:1997a}, 
we use $\delcg=1.5$ for the model with massive neutrinos and $\delcg=1.3$ for
all other models, in this figure (but cf. Table~\ref{tbl:psvirnorm} for
best-fit $\delcg$ and $\delct$ to our simulation results).  The observational
cluster abundance estimates plotted are in reasonable agreement with these mass
functions, especially if the mass estimates are low as indicated by some
gravitational lensing estimates.

In figure~\ref{fig:cmb}, we compare each model to several
recent CMB measurements, using the \textsc{cmbfast} program of
\citeN{seljak:1996}.
\begin{figure*}
\ifref\else
\begin{figwide}{cmbcmp}
\fi
  \caption[Model Comparison to Cosmic Microwave Background.]
          {Model Comparison to Cosmic Microwave Background.  All models except
           SCDM are consistent with \cobe\ four year data
           (\protect\citeNP{gorski:1996a}; G\'orski, private
           communication).
           Circles, solid squares, open squares and asterisks are the
           \cobe\ four year power spectrum \protect\cite{tegmark:1996}, Saskatoon
           1995 results \protect\cite{netterfield:1997}, CAT detection
           \protect\cite{scott:1996} and Python III results
           \protect\cite{platt:1997}, respectively.  Not shown are systematic
           normalization errors of 14 and 20 per cent, for Saskatoon and Python
           III, respectively.  The curves are all calculated using
           the \textsc{cmbfast} program of \protect\citeN{seljak:1996}.
           Cosmological parameters correspond to models considered in this
           paper, except for SCDM.  The normalization is adjusted so that the
           low harmonics match the output of our linear code. \textsc{cmbfast}
           is capable of calculating larger multipoles than our linear code. 
	   SCDM is shown here with
           $\omb=0.1$, since all the high-$\ell$ features in the CMB spectrum
           are dependent upon baryon interactions, but was actually simulated
           with no baryons.}
  \label{fig:cmb}
\ifref\else
\end{figwide}
\fi
\end{figure*}
We also show the four most recently announced CMB results on the figure.  Not
shown are systematic calibration errors of 14 per cent for Saskatoon and 20 per
cent for Python III.  Note that the OCDM model is strongly inconsistent with
the Saskatoon points, and our choice of $\Omega_0=0.5$ is at the 95 per cent
confidence lower limit for an open model \cite{lineweaver:1997b}. Also note
that the models with even the relatively mild tilt of $n=0.9$ are at best
in marginal agreement with the Saskatoon data around the first Doppler peak.

\begin{figure}
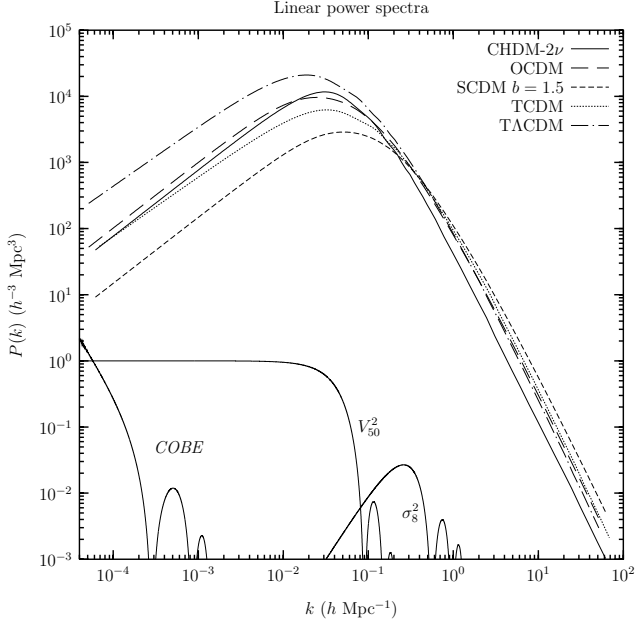

\ifref\else
\begin{fig}{power_lin}
\fi
  \caption[Linear power spectra]
          {Linear power spectra used in our simulation suites.  Also shown are
           two of the window functions used in normalizing the models:
           $k^2W^2(rk)$ with $r=8$ $\hmpc$ for $\sigma_8^2$
           and $W^2(rk)$ with $r=50$ $\hmpc$ for $V_{50}^2$.  Here
           $W(x)=3\left[\sin(x)-x\cos(x)\right]x^{-3}$.  Also shown (for
           illustrative purposes only) is the
           equivalent window function for approximate \cobe\ normalization
           using the pure Sachs-Wolfe effect,
           $j_{10}^2(d_{\mathrm{h}}k)/(2\pi d_{\mathrm{h}}^2k^2)$ where
           $j_{10}(x)$ is the 10th order spherical Bessel
	   function and $d_{\mathrm{h}}$ is the horizon
           distance.  The version plotted has the amplitude raised by a factor
           of 100 for visibility and uses $d_{\mathrm{h}}=2c/H_0=6000$
           $\hmpc$, which is appropriate for $\Omega_0=1$.  For OCDM, the
           horizon distance is 7470 $\hmpc$ and for T$\Lambda$CDM, it is
           8810 $\hmpc$, so the window function moves a small distance to
           smaller $k$ in those cases.  A similar window
           function for cluster abundance doesn't exist because it doesn't
           have the form of a convolution.  In an extremely rough sense, the
           scales are comparable to those sampled by $\sigma_8$.}
  \label{fig:linpow}
\ifref\else
\end{fig}
\fi
\end{figure}

\begin{figure}
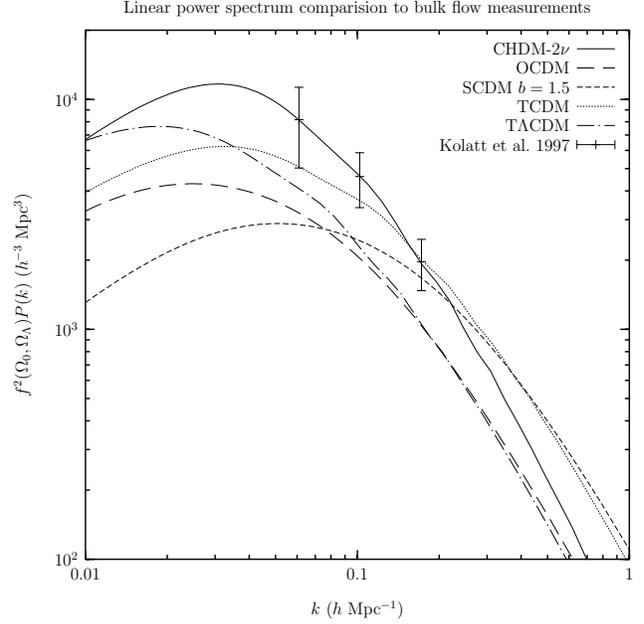

\ifref\else
\begin{fig}{power_bf}
\fi
  \caption[Linear power spectrum comparison to bulk-flow measurements]
          {Linear power spectrum comparison to bulk-flow measurements.  The
           curves are all a magnification of figure~\ref{fig:linpow}, multiplied
           by $f^2(\Omega_0,\oml)\equiv (a\dot{D}/\dot{a}D)^2$.  The
           three data points are from \protect\citeN{kolatt:1997}.
           $f(\Omega_0,\oml)$ was calculated exactly, using equation
           (C.3.14) of \protect\citeN{gross:thesis} and its analytic
           derivative.}
  \label{fig:bfpow}
\ifref\else
\end{fig}
\fi
\end{figure}

Figure \ref{fig:linpow}%
shows the linear power spectra at the present epoch.  As one might expect, all
the spectra nearly cross at a wavenumber of a few tenths $\hmpcinv$,
corresponding to cluster scales.  Also, we show some of the window functions
used in the normalization procedure described above.

In figure~\ref{fig:bfpow}, we compare our models to the matter power spectrum
recently measured from bulk flows by \citeN{kolatt:1997}.
We only use the three data points that Kolatt \& Dekel use for their own
statistical analysis, because for larger wavenumbers smoothing lowers the power
significantly.  SCDM disagrees at about the 2.5$\sigma$ level, also reflected
in its low value of $V_{50}$ in table~\ref{tbl:linear}.  OCDM and T$\Lambda$CDM
disagree because the value of $P(\sim 0.1$ $\hmpcinv)$ is fixed by
comparing the observed density of clusters (WEF93,
BGGMM93, \citeNP{borgani:1997b})\nocite{biviano:1993,white:1993} to
the Press-Schechter prediction, and they have low values of
$f(\Omega_0,\oml)\equiv\dot{D}a/D\dot{a}\approx\Omega_0^{0.6}$,
where $D(\Omega_0,\Lambda,t)$ is the linear growth factor and $a(t)$ is the
expansion parameter.  The combination of cluster abundances and bulk-flow power
spectrum measurements favors $f\sim 1$, for the currently favored classes of
CDM-variant models.

There is currently significant controversy over the proper normalization of
models, and our OCDM and T$\Lambda$CDM normalizations are higher than the
recent fits reported in \citeN{eke:1996}, based on cluster
X-ray temperature distributions \cite{henry:1991},%
\footnote{Note that \protect\citeN{eke:1996} discovered two
compensating errors in the \protect\citeN{henry:1991} analysis: an arithmetical
error of a factor of 4.2 and a binning error of a factor of about 4 in the
other direction.  The errors have also been noted in
\protect\citeN{viana:1996}.}
though they are
consistent with the older analyses of WEF93\nocite{white:1993} and the
newer cluster velocity dispersion measurements of \citeN{borgani:1997b}.
\citeN{pen:1996} has reanalyzed the \citeN{eke:1996} calculation, and he
gets slightly higher low-$\Omega_0$ normalizations of $\sigma_8=0.86$ and
$0.72$ for our T$\Lambda$CDM and OCDM models, respectively.
These normalizations are close to those we have chosen
(table~\ref{tbl:linear}).

\subsection{Algorithm}
\label{sec:algorithm}

A classic problem with gravitational simulations is the `overmerging'
problem, where small scale structure in highly overdense regions is not
resolved.  Part of the problem is physical -- real galaxies form much denser
cores than dissipationless halos can, because the baryons can dissipate energy
(but cf. \citeNP{klypin:1997}).
Aside from that, numerical limitations can make the problem vastly worse.
There are two numerical effects to consider:
force resolution and sampling of initial conditions and bound structures.
Improving either of these requires vast amounts of memory and processing time,
so there is an inherent tradeoff.

Recently, the more popular approach has been to improve the forces by using
hybrid \cite[for example]{hockney:1988,couchman:1991,xu:1995} or adaptive-mesh
\cite[for example]{kravtsov:1997} force solvers, at
the expense of either poor sampling of initial fluctuations or small box
sizes.  We choose a complementary approach, where we try to balance the
sampling of density in a large box with the force resolution.  We still
require a large dynamic range in order to sample small scales well and
simultaneously simulate a large volume for comparison to redshift surveys.
Since the two requirements imply an enormous number of particles, computer time
limitations force us to use the \emph{fastest} code available.  We choose a
standard particle-mesh \cite{hockney:1988}
algorithm, parallelized to run on a distributed-memory message-passing system.%
\footnote{Specifically, the Cornell Theory Center SP2, but the code is portable
to any system supporting MPI, including heterogeneous workstation clusters and
most modern supercomputers.}
This type of code produces adequate forces at about 1.5 grid cells
(KNP97, Appendix A)\nocite{KNP:1997}, but we
double this distance to be conservative.  So, we require that we have $3^3$
times as many grid cells as particles, for the high resolution suite.  We
choose a grid cell size of 65 $\hkpc$, with $\Ng=1152^3$ grid cells and
$\Np=384^3=57$ million `cold'
particles.  For the large-volume case, we wish to follow the dynamics only
of clusters of galaxies, so we can afford to coarsen the density grid slightly.
We find that a cell size of $390$ $\hkpc$ is adequate for following the
dynamics of $\ga 10^{14}$ $\hmsun$ objects, and expect information about
smaller objects to come from the high resolution simulations.  The slight
coarsening of the density resulted in a substantial advantage in running time.

Initial conditions were calculated using a
parallelized \citeN{zeldovich:1970} approximation.  For CHDM models, we started
with a uniform grid of cold particles, and two neutrinos at the position of
every cold particle.  Cold particles and neutrinos were offset from the grid
using separate cold+baryon and hot power spectra, and consistent velocities were
derived from the offsets using scale-dependent linear growth rates
calculated by a refinement of the
\citeN{holtzman:1989} code.  In addition, equal and opposite random
thermal velocities were chosen for each pair of neutrinos from a redshifted
relativistic Fermi-Dirac distribution \cite{KHPR:1993}.

We adopted the form of the equations of motion used in
\citeN{kates:1991} generalized to arbitrary cosmology:
\begin{equation}
  \nabla^2\phi=\frac{3}{2a}\delta%
          \equiv\frac{3}{2a}\frac{\delta\rho}{\Omega\rho_{\mathrm{c}}},
\end{equation}
\begin{equation}
  \frac{{\mathrm{d}}\bvec{p}_{\mathrm{i}}}{{\mathrm{d}}a}=%
                 - \dot{a}\bvec{\nabla}{\phi}(\bvec{x}_{\mathrm{i}}),
\end{equation}
and
\begin{equation}
  \frac{{\mathrm{d}}\bvec{x}_{\mathrm{i}}}{{\mathrm{d}}a}=%
                 \frac{\bvec{p}_{\mathrm{i}}}{a^2\dot{a}},
\end{equation}
where $\dot{a}$ is given by the Friedmann equation with time variable $H_0t$,
\begin{equation}
\dot{a} = \adot.
\end{equation}
Time discretization was a standard `leapfrog' scheme (cf.
\citeNP{hockney:1988}), with even steps in the expansion parameter $a$.
To reduce the expense of the simulations, the timestep was chosen only to
stabilize bound structures at the final timestep, rather than keep \emph{all}
structures on the scale of the grid spacing stable.  This is only a problem
for the cores of clusters, which have the highest velocities.
For clusters, we presume an upper bound of particle velocities of 1200 $\kms$
today and a minimum diameter of any given bound structure equal to
the linear cell size.  Stability for such an object requires that particles
take at least one timestep to traverse the object.  So, the required condition
is
\begin{equation}
  \Delta a\equiv \dot{a}\Delta t\la \frac{H_0L}
       {N_{\mathrm{g}}^{1/3}v_{\mathrm{max}}}
\end{equation}
independent of cosmology because the condition is evaluated at the present
epoch and $H_0L$ is chosen to be the same for all models.  Plugging in
$v_{\mathrm{max}}=1200$ $\kms$, $H_0L=7500$ $\kms$ and
$N_{\mathrm{g}}=1152^3$ gives $\Delta a\la 0.005$, or 200 timesteps for the
high resolution suite.  Such a low $v_{\mathrm{max}}$ will not model the
interiors of clusters well, since they are observed to have velocity dispersions
larger than that, but to remain bound to the cluster, particles have the
much looser requirement that they not traverse the \emph{whole cluster} in one
timestep.  As large cluster radii are up to about 50 grid cells, the effective
stability limit is 20 per cent the speed of light inside a large cluster, for
the high resolution suite, presuming that the cluster is adequately modeled by
an isothermal sphere.  We checked that the choice of timestep was adequate by
running a 25 $\hmpc$ box CHDM-2$\nu$ simulation with 384$^3$ grid cells
(which has the same 65 $\hkpc$ cell size as the high resolution suite) for
200 timesteps and for 300 timesteps.  The resulting mass functions were not
significantly different.  For the large volume suite, clusters do not cover
nearly as many cells as in the high resolution suite, and so the velocity limit
is much higher.  Our choice of 150 timesteps corresponds to a limiting speed of
5000 $\kms$ if particles are not to cross one cell in a timestep.  The
suite parameters are summarized in table~\ref{tbl:suites}.%
\footnote{Though the implementation, and especially parallelization, of the
two-species particle mesh code described above is much less trivial than one
might suppose, discussion of the code has been omitted for space
considerations.  The interested reader may find a detailed description of the
code, its implementation on the Cornell SP2, and several code tests in
\protect\citeN{gross:thesis}.}
\begin{table*}
  \tblcapt{
    \begin{tabular}{ccccccccc}
    Suite & Box size & $N_{\mathrm{cells}}$ & Cell size & $N_{\mathrm{cold}}^a$ & $M_{\mathrm{cold}}^b$ & $N_{\mathrm{steps}}$ & $M_{\mathrm{min}}^c$ & $f^d$\\
          & $\hmpc$ && $\hkpc$ && $(\omc+\omb)$ $\hmsun$ &&$\Omega_0$ $\hmsun$ &\\
    \hline
    high res &  75 & 1152$^3$ &  65 & 384$^3$ & $2.09\times 10^9$    & 200 & $3.4\times 10^{11}$ &0.078\\
    low res  & 300 &  768$^3$ & 390 & 384$^3$ & $1.34\times 10^{11}$ & 150 & $7.3\times 10^{13}$ &0.043\\
    \hline
    \end{tabular}
  }
  {Simulation parameters for both simulation suites}
  {tbl:suites}
  {$^a$ Number of cold particles; for models with massive neutrinos,
        $N_{\mathrm{hot}} = 2N_{\mathrm{cold}}$.\\
   $^b$ Mass of cold particles; for models with massive neutrinos,
        $M_{\mathrm{hot}} = M_{\mathrm{cold}}\omh / 2(\omc +\omb)$.\\
   $^c$ Halo detection cutoff, from the restriction that halos must be
        larger than the grid size.\\
   $^d$ Fractional error in mass for the smallest halos identifiable.\\
  }

\end{table*}

The \citeN{zeldovich:1970} approximation is only valid when the
rms fluctuations are much less than 1.  In practice, one picks a starting time
early enough so that linear theory brings the rms fluctuations well below 1.
The initial time was chosen so that the rms overdensity on the grid scale was
$\delta_{\mathrm{rms}}\la 0.2$.  This was $z=30$--$60$, depending on
the model.  Particle data and halo catalogs were stored at four equally
spaced intervals in $a$ during execution.  The large volume simulation suite
used the same starting times as the high resolution suite even though they
could have been started somewhat later due to the poorer resolution.  The extra
computation involved is about one timestep and is therefore negligible.

Random numbers are necessary to model inflation-generated Gaussian fluctuations
and random phases in the density field.  Such randomness introduces highly
significant variation from simulation to simulation, commonly referred to as
`cosmic variance.'  Because we can only observe one universe, quantifying
the effect of cosmic variance is very important and is a separate issue from
variations between models due to different physics.  In these suites, we have
separated the effects by picking a single random number seed for each suite,
checking that the largest 26 waves do not have any fluctuations larger than
a factor of 2, and rerunning one model with a different seed.  That is, within
each suite, the random numbers for each model within a suite are all the same,
and large wavelength fluctuations are restricted to a smaller range than
Gaussian statistics would permit, in an attempt to prevent rare statistical
flukes from compromising expensive simulations (as happened in
\citeNP{KHPR:1993}).
This means the structures are approximately in the same place, and when one
also considers the cluster abundance criterion discussed in
section~\ref{sec:models} there is roughly the same number, distribution and
positions of $5\times 10^{14}$ $\hmsun$ clusters in all the models in a given
suite.  Note that the models do have different power spectra and fluctuation
growth rates, so distributions can differ for objects with different masses.

\section{Dark matter halos}
\label{sec:halos}

\subsection{Halo finding algorithm}
We identify dark matter halos using a spherical overdensity algorithm similar
to that of KNP97\nocite{KNP:1997}, with some of the limitations removed:
\begin{enumerate}
\item We define candidate halos as the centers of all density maxima containing
an overdensity greater than
$\delta\equiv\delta\rho/\Omega_0\rho_{\mathrm{c}}=50$.  A density
maximum is defined as a cell whose density is greater than its six Cartesian
neighbors.  Just in case there are other halos hiding in those six neighbors,
we also consider each of them to be candidates.  Note that the finite grid size
(as in all other grid-based halo finders, such as \textsc{denmax},
\citeNP{gelb:1994a}) will introduce a minimum separation between
halos, which may cause small halos to be missed, which in turn will require a
mass cut.
\item Each candidate halo then has the location of its center set iteratively
to the center of mass of all the particles inside a sphere of diameter equal
to the cell size (65 $\hkpc$ in our high resolution case).  Halos are
expected to have a minimum size of the order of the grid size, so this procedure
moves the candidate halo to the peak of the density maximum.  Of course, since
we have defined more than one candidate for each detected maximum, some
candidates will converge on the same halo.  The smaller mass object in a given
pair is removed if the distance between the centers of mass is less than half
the grid spacing.
\item We perform a central overdensity cut.  All halos that don't enclose a
mean overdensity sufficient for virialization according to the spherical
collapse model (see \citeNP[appendix C]{gross:thesis}, and references therein)
at the end of the center-of-mass detection phase are presumed
not to be virialized objects and are discarded.  Typically, this reduces the
number of halos by a factor of 2--3, though the number is model dependent.
\item We now estimate at what radius the mean enclosed overdensity
$\delta\equiv\delta\rho/\Omega\rhocrit$ falls to $\dvir$, the virial radius of
the halo in spherical infall models.%
\footnote{Note that our definition of $\dvir$ is related to
\protect\citeN{eke:1996} by $1+\dvir=\Delta_{\mathrm{ECF}}/\Omega$.
Our choice is appropriate for the density field calculations in an $N$-body
code.}
For each halo, we count the number of particles within five radii up
to five grid cells (325 $\hkpc$ in this case) away from the center,
convert that to density, and interpolate the radius at which
$\delta=\dvir$ ($\rvir$) using power-law cubic splines.  If five radii is not
large enough to enclose $\rvir$, we search five more radii, each twice as
long as the original radii.  This is repeated until we enclose $\rvir$.
\item We define the mass of the halo as the mass enclosed in $\rvir$.  The
velocity is the mean velocity of all the particles within $\rvir$.
\item In general, the largest halos in a high resolution run contain much
resolved but bound substructure.  Because we search for the $\delta=\dvir$
radius, we detect the same regions of space dozens of times for the largest
halos.  To remove `double-counted' halos, the halos are searched in reverse
order by mass to see if they enclose the centers of any smaller halos.  If so,
the smaller halo is thrown away.  Note that the ordering is important because
three-body intersections would be non-deterministic otherwise, and throwing
away halos that only intersect is too stringent.
\end{enumerate}

One limitation of this algorithm is that it presumes all halos are
spherically symmetric, which is demonstrably untrue.  But the effect on
the mass function is \emph{random}, rather than systematic,
and finding the halos with an algorithm generalized to \emph{ellipsoidal}
distributions does not change the mass function significantly, even though it
changes the parameters of individual halos.
Because the halos have finite size, one cannot perform mass-weighted correlation
function analyses, for distances less than the largest halo radius
(about 2--3 $\hmpc$ in radius, typically).

The other limitation is the use of the density grid to identify halo candidates.
If one considers a worst-case identification where a large number of particles
all collect in one corner of a grid cell, in order to guarantee that all nearby
halos are identified, one must draw a sphere which encloses the \emph{entire}
cell, of radius
\begin{equation}
\rmin = \sqrt{3}L/\Ng^{1/3},
\label{eqn:rmin}
\end{equation}
where $L$ is the length of one side of the computational volume and
$\Ng$ is the number of grid cells.  If halos happen to
be bigger than that, then the last step of the halo catalog generator makes it
unimportant that we couldn't see nearby structure.  Fortunately, halo extent
is trivially related to halo mass because we have defined both where the
mean overdensity is $\delta=\dvir$.

\subsection{The effect of mass resolution}

To what extent should you, the reader, trust the mass functions
presented in this paper?  To answer that, one must consider several effects.
A typical feature in a mass function is that the large-mass end
becomes `wiggly,' usually blamed on the scarcity of high mass halos
combined with cosmic variance.  There is a related effect at somewhat
smaller masses, since very large halos tend to have somewhat massive companions.
For example, in most models in our high resolution suite, $5\times 10^{13}$
$\hmsun$ objects are fairly rare, but it is common to see them as
companions for $10^{15}$ $\hmsun$ objects.  So, the wiggles may
propagate down the mass function, and cosmic variance may have a significant
effect on more than just the highest mass scales.

Cosmic variance fortunately leaves a signature, in that the mass function is
not smooth at high masses.  But, it is quite important to figure out the
limiting factors at low mass, where typical mass functions are quite smooth.
What limits accuracy here are the effects of finite sized grids and finite
numbers of particles.

The effect of the finite sized grid in identifying maxima in the final particle
distribution was discussed above, and one must merely translate the minimum
radius of a halo $\rmin$ to a minimum mass.  Since the halo radius and mass are
defined as enclosing a mean overdensity of $\dvir$, the mass $\Mvir$ of a halo
of radius $\rvir$ is
\begin{equation}
\Mvir = (1+\dvir)\frac{4\upi}{3}\Omega_0\rhocrit\rvir^3.
\end{equation}
So, a \emph{very} conservative mass cut is
\begin{equation}
\mmin = (1+\dvir)\frac{4\upi}{3}\Omega_0%
              \rhocrit\frac{(L\sqrt{3})^3}{\Ng}.
\label{eqn:mmin}
\end{equation}
Plugging in values for the high resolution suite, the mass cut is
$3.4\times 10^{11} \Omega_0$ $\hmsun$.  For simplicity, we make the
same mass cut on all models, corresponding to $\Omega_0=1$.

One might worry that the central density cut described in the previous
section could cut too many small halos, because the fairly long timesteps
used cause the density within the `half-mass' radius to go down
by about a factor of two if the timestep equals the stability limit
\cite{quinn:1997}.  We perform the central overdensity cut at $r=L/2\Ng$,
but the proximity restriction used in deriving equation (\ref{eqn:mmin})
requires that
halo radii in the final catalogs be at least $\sqrt{3}L/\Ng$.  If halo profiles
fall at least as fast as $r^{-1}$ (whereas the \citeNP{navarro:1996b} profile
says it should be much steeper than that near the virial radius), then the
density fed into the central overdensity cut should be at least a factor of
$2\sqrt{3}\approx 3.4$ greater than the virial density.  This more than offsets
the density smoothing due to timestepping at the stability limit, so we neglect
the effect of time steps in our mass resolution analysis.  Note that our
timesteps are only near the stability limit for virial radii near the detection
limit -- otherwise, a particle takes many timesteps to cross a halo.
Therefore, lowered densities due to long timesteps are only a concern for the
smallest detectable halos.

One might also worry that the quality of the force law at scales approaching
the grid scale would also result in reduced central density.  At the 1.7 grid
cells proximity cutoff, the point-mass potential in our simulations is about 90
per cent of the correct $GM/r$ value.  With such a force law, the virial
theorem requires that the density be also 10 per cent low, to maintain the same
velocity dispersion.  Thus, some of the smallest halos around the mass cut will
not make it into the catalog.  In practice, the density profiles for the
smallest halos are considerably noisier than 10 per cent due to asphericity and
background particles, so we neglect the effect of an oversoftened force law.

Particle discreteness may also affect the halo mass function, because
random fluctuations may affect the detection of some of the smallest halos.
We consider here how much significance we need to make the expected number of
halos missed fewer than the number of halos in the simulation.
Because halo boundaries are defined where the
mean overdensity $\delta$ is $\dvir$ and the mass of a particle is
\begin{equation}
M_{\mathrm{p}}=\Omega_0\rhocrit \frac{L^3}{\Np},
\end{equation}
where $\Np$ is the number of cold particles in the simulation,%
\footnote{For simplicity, we consider the significance of models with only
one particle mass.  For CHDM-2$\nu$, a given mass will always be represented
by more particles than in SCDM.  So, one can get a conservative estimate of
the significance by only considering the cold particles.}
the number of
particles inside a halo of mass $\Mvir$ in a simulation box of size $L$ is
\begin{equation}
\Nvir = \frac{\Mvir\Np}{\Omega_0\rhocrit L^3}.
\end{equation}
For counting $N$ particles within $r$, the random variation in number is
$\sigma=\sqrt{N}$.  Let us suppose there are $N_{\mathrm{h}}$ halos above a
given mass, and we wish to detect them all.  We presume that counting halos
is a Gaussian process and state that the $n$-sigma uncertainty in the detection
of the halos corresponds to incorrectly detecting or missing a fraction
$\textrm{erfc}(n/\sqrt{2})$ of the halos.  We require detection of all halos, so
the fraction missed should be less than $1/\rhoh L^3$, where $\rho$ is the
number density of halos above the mass cutoff, and $L^3$ is the volume of the
simulation box where halos are identified.  Inverting, we need detections of
$\sqrt{2}\,\textrm{erfc}^{-1}\left(1/\rhoh L^3\right)$ sigma.
The density $\rhoh$ should really come from the Press-Schechter approximation,
given a desired mass cutoff, but the inverse complementary error function is
extremely insensitive to the value of its argument, once it becomes much less
than one.
As an example, for $\rhoh=1$ $h^3$ Mpc$^{-3}$ (appropriate for a mass cutoff a
little below $10^{11}$ $\hmsun$ for most models), we need to have at least
4.7$\sigma$ detections of all halos. Less significant detections mean it is
likely some of them have been missed by random fluctuations.
This means every halo must contain at least 23 particles.
More generally,
\begin{equation}
\nmin = 2\left[\textrm{erfc}^{-1}\left(\frac{1}{\rhoh L^3}\right)\right]^2
\end{equation}
for the rather liberal restriction that we only require detection of the halo.

\begin{figure}
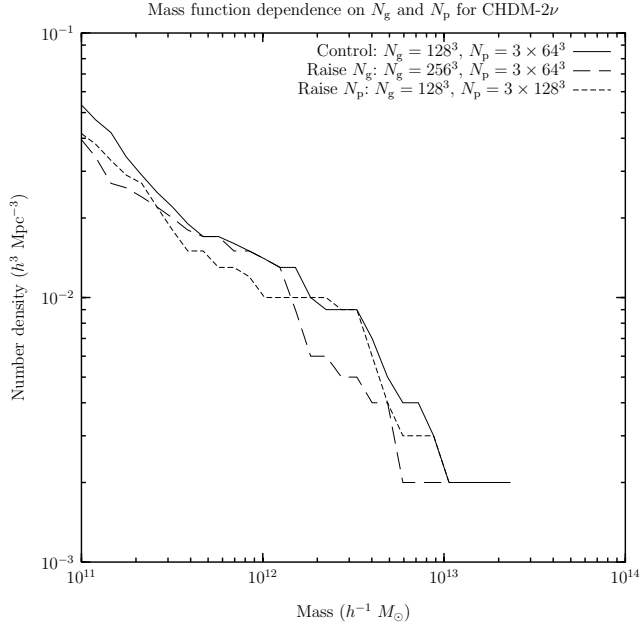

\ifref\else
\begin{fig}{restest}
\fi
  \caption[Mass and force resolution tests]
          {Effect of raising the number of particles or the number of grid
           cells by a factor of 8 in a very small CHDM-2$\nu$ 10 $\hmpc$
           simulation with $\Ng=128^3$ and $\Np=3\times 64^3$.
           The mass functions are not
           significantly different.  A somewhat low mass cutoff of $10^{11}$
           $\hmsun$ has been applied.  The high resolution suite
           has a linear cell size that is slightly smaller than the
           $\Ng=128^3$ runs shown in this figure.}
  \label{fig:restest}
\ifref\else
\end{fig}
\fi
\end{figure}

As an alternative cutoff criterion, requiring a 10 per cent or less 1-$\sigma$
error in mass is a more stringent requirement, and every halo must have at
least 100 particles, since mass is determined by counting particles within
several radii.  If one requires a fractional error of $f$ for a minimum
halo mass of $\mmin$, one needs at least
\begin{equation}
\Np = \frac{\Omega_0\rhocrit L^3}{f^2 \mmin}
\end{equation}
particles in the simulation.
The parameters used, and the effective $f$ they
allow, are shown in table~\ref{tbl:suites}.
Figure \ref{fig:restest}%
shows that, with grid sizes and mean interparticle
spacings of the order of those used in our suite, the effect of lowering
either the grid size or the mean interparticle spacing by a factor of two
does not significantly affect the mass function.  For this test, we raised
the threshold for halo candidate identification from $\delta=50$ in one cell
to $\delta=70$ because one isolated cold particle in the high $\Ng$
case gives $1+\delta=51.2$.

\begin{figure}
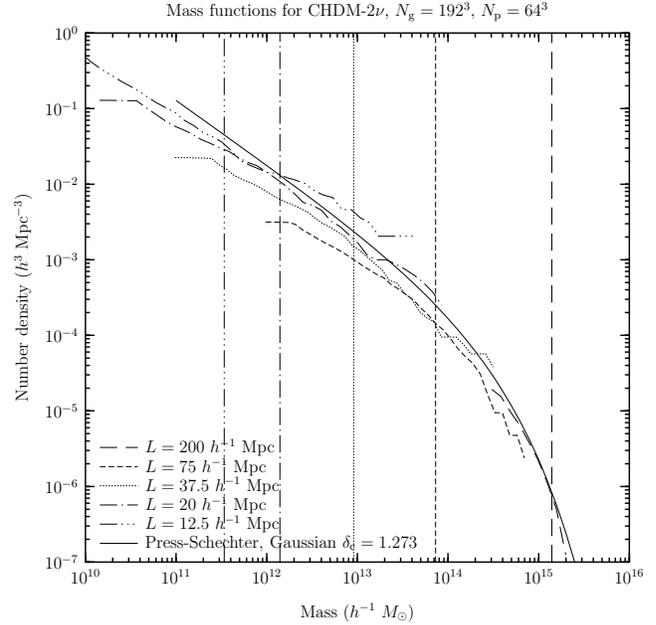

\ifref\else
\begin{fig}{mftest}
\fi
  \caption[Effect of grid size on mass functions]
          {Effect of grid size on mass functions.  The curves represent very
           small simulations of various sizes.  The Press-Schechter model was
           tuned to match the cluster-scale part of the mass function in the
           largest box.  Note that the envelope mirrors the Press-Schechter
           curve reasonably well, but each individual mass function has a
           power-law index that is too shallow.  Mass functions are limited at
           the large-mass end by statistics -- one simply runs out of enough
           space to create objects in -- and on the small-mass end by some
           fraction of the halos becoming as small as two grid cells, which
           means it is not guaranteed that the halo can be resolved from its
           neighbors, particularly if they are also small halos.  The vertical
           lines represent the lower limit in mass for each run, above which
           all halos can be detected.}
  \label{fig:mftest}
\ifref\else
\end{fig}
\fi
\end{figure}

To explicitly test the effect of grid sizes on our mass functions, we ran five
small simulations of the CHDM-2$\nu$ model with $\Ng=192^3$ grid cells and
$\Np=3\times 64^3$ particles, with various-sized boxes.  Though these
simulations are too small to generate meaningful mass functions on their own,
collectively their upper envelope does match the Press-Schechter formula
reasonably well, for $\delcg=1.2$ with a Gaussian filter.
Figure~\ref{fig:mftest}%
shows the five different mass functions.  Also shown are lower mass cuts,
determined for every model using equation~(\ref{eqn:mmin}).  Above the mass
cutoffs, every mass function agrees with the one for the next smaller box.
Well below, the mass function slopes are not steep enough, but they agree with
the neighboring curves for significant distances below the mass cuts, so
it may be reasonable to extrapolate the mass function further.  Every halo
detected by the halo finder is represented in the figure, and the locations of
the lower mass cuts are indicated by vertical lines.  This test could
conceivably overproduce clusters because of the extremely poor force and mass
resolutions in the largest volume run -- a cell width is about the size of an
Abell radius.  This $\delc$ result does persist for much larger simulations, as
discussed below.

\section{Results}
\label{sec:results}
\begin{figure}
\ifref\else
{\Large\resizebox*{\linewidth}{!}{{
\setlength{\unitlength}{0.1bp}
\begin{picture}(4320,1512)(0,0)
\special{psfile=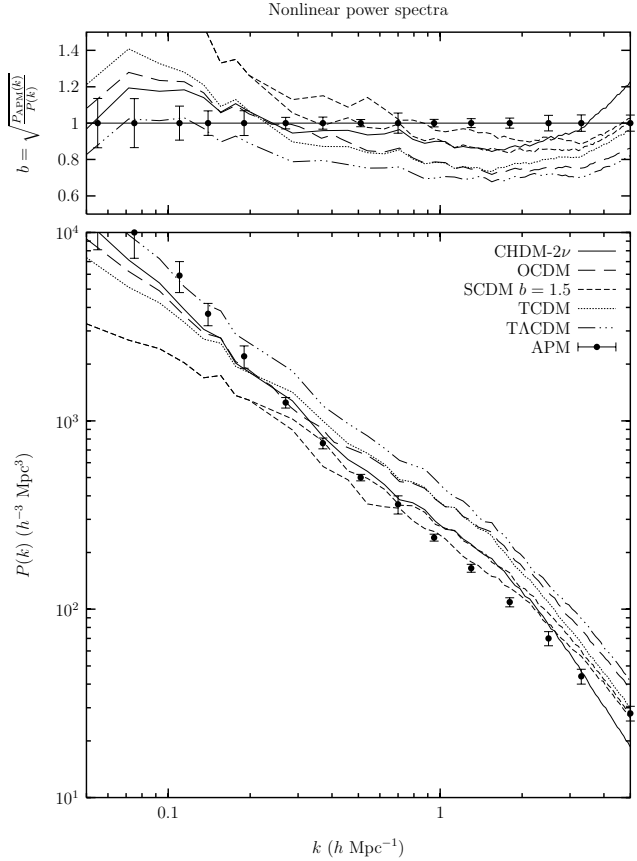 llx=0 lly=0 urx=864 ury=353 rwi=8640}
\put(2385,1412){\makebox(0,0){Nonlinear power spectra}}
\put(100,631){%
\special{ps: gsave currentpoint currentpoint translate
270 rotate neg exch neg exch translate}%
\makebox(0,0)[b]{\shortstack{$b=\sqrt{\frac{P_{\mathrm{APM}}(k)}{P(k)}}$}}%
\special{ps: currentpoint grestore moveto}%
}
\put(450,1136){\makebox(0,0)[r]{1.4}}
\put(450,883){\makebox(0,0)[r]{1.2}}
\put(450,631){\makebox(0,0)[r]{1}}
\put(450,379){\makebox(0,0)[r]{0.8}}
\put(450,126){\makebox(0,0)[r]{0.6}}
\end{picture} }}}
\begin{fig}{power_nl}
\fi
  \caption[Nonlinear real-space dark matter power spectra]
          {Nonlinear real-space dark matter power spectra, compared to the APM
           real-space galaxy power spectrum \protect\cite{baugh:1994} of
           galaxy number-count fluctuations.  The simulation power spectra shown
           here are a composite of the high and low resolution suites, where
           data from a model's high resolution run is used at large $k$ and
           low resolution data is used at small $k$.  Two different high
           resolution runs of the SCDM case are shown as a guide to how large
           cosmic variance is.  The power in the second SCDM realization is
           20--30 per cent lower than that in the first realization for $0.3\la
           k\la 1$ $\hmpcinv$.  The APM data are presumably biased with
           respect to the matter power spectrum, and yet the OCDM, TCDM,
           and T$\Lambda$CDM cases require the APM data to be significantly
           \emph{antibiased} with respect to the dark matter, with $b^2\sim 0.6$
           for OCDM and TCDM and $b^2\sim 0.5$ for T$\Lambda$CDM at $k\sim 1$
           $\hmpcinv$.  If APM misses galaxies in clustered regions, that
           would give a low power spectrum on scales of $k\ga 1$ $\hmpcinv$
           (see text).}
  \label{fig:nlpow}
\ifref\else
\end{fig}
\fi
\end{figure}

The connection between simulations and observations is still fairly uncertain,
and the least well determined portion of it is the galaxy
identification procedure.  It is therefore helpful to do as much analysis as
one can using quantities that are insensitive to the details of galaxy
formation.  Currently, only bulk flow motions \cite{kolatt:1997} provide a
meaningful \emph{matter} power spectrum, but the large smoothing required means
that the comparison is best made to the linear power spectrum (see
figure~\ref{fig:bfpow}).  When investigating quantities derived from
observations of galaxies (as the vast majority of astronomical observations
are) one is forced to make assumptions based on expections about the nature of
galaxy bias, for example the usual expectation that galaxies are more clustered
than the dark matter.  Figure \ref{fig:nlpow}%
shows nonlinear real-space dark matter power spectra for
all our models, compared to the APM real-space galaxy power spectrum
\cite{baugh:1994}.  The OCDM model requires significant antibiasing and the
$\Lambda$CDM model requires even more.  There is no evidence for such strong
antibiasing, and it is very difficult to explain physically, especially on such
large scales (cf. \citeNP{yepes:1997,kauffmann:1997}).  Additional
arguments against strongly scale-dependent antibiasing are given in
KPH96\nocite{klypin:1996a}.

\begin{figure}
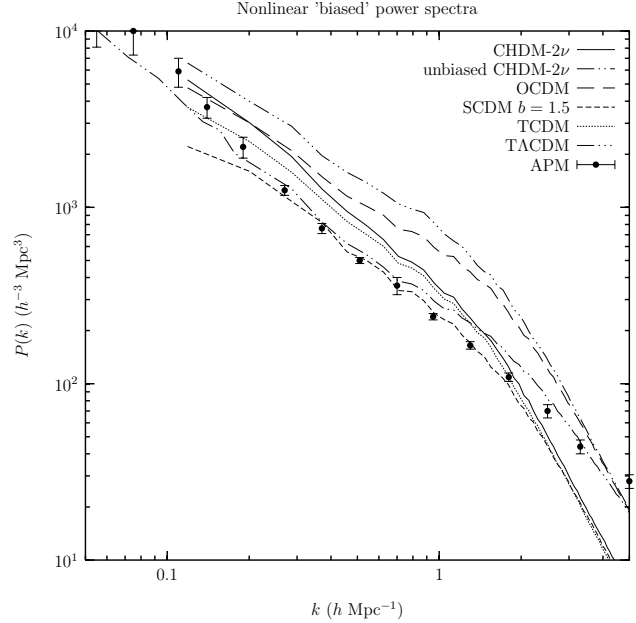

\ifref\else
\begin{fig}{power_bw}
\fi
  \caption[Nonlinear biased dark matter power spectra]
          {Nonlinear power spectra, presuming an extreme scale-dependent
           biasing scheme.  The density field has been set to `on' at any cell
           containing mass exceeding the largest particle mass in the 75
           $\hmpc$ suite, $2.1\times 10^{9}$
           $\hmsun$, and `off' everywhere else.  That mass cut
           is most likely lower than anything that could make it into the CfA2
           or APM catalogs, except if one assumes an impossibly small
           mass-to-light ratio.  The result of such a bizarre galaxy
           identification scheme is a bias on large scales, due to clearing
           out the void regions, and an antibias on small scales, due to
           removing the high peaks in density.
           We do comparisons with the high resolution suite
           because the low resolution suite particle mass is too high.}
  \label{fig:bwpow}
\ifref\else
\end{fig}
\fi
\end{figure}

The process of galaxy formation is not well understood, so one could argue
that perhaps there \emph{is} some mechanism that would give us strong
antibiasing.  We have created an extreme model for galaxy formation designed to
produce as much antibias as possible (cf.  KPH96\nocite{klypin:1996a}).
Everywhere in the density grid, if there is more than $2.1\times 10^{9}$
$\hmsun$ in a grid cell, we presume one galaxy forms there.  That mass
corresponds to slightly more than the mass due to one isolated particle in the
high-resolution SCDM and TCDM simulations (which have the most massive
particles in the suite).  Such a limit is necessary to prevent placing excess
power in the voids due to vestiges of the initial grid there.  This is a highly
unreasonable model for galaxy formation, as it says that the density of
$\ga\!\!2\times 10^{11}$ $\hmsun$ galaxies in the core of the Coma cluster
should be the same as in the local group, and this is clearly ruled out
observationally.  However, even though there is significant antibias on small
scales, it is only visible at scales smaller than about $k=1$ $\hmpcinv$ (see
figure~\ref{fig:bwpow}),%
whereas antibiasing is needed on scales larger than that in order for OCDM
or T$\Lambda$CDM to be consistent with the APM power spectrum.  Note that a 
possible way out of the antibiasing requirement is to note that the APM survey
is incomplete in clustered regions, which will raise the `true' power spectrum
above the APM measurement on small scales (Zabludoff, private communication).

\begin{figure}
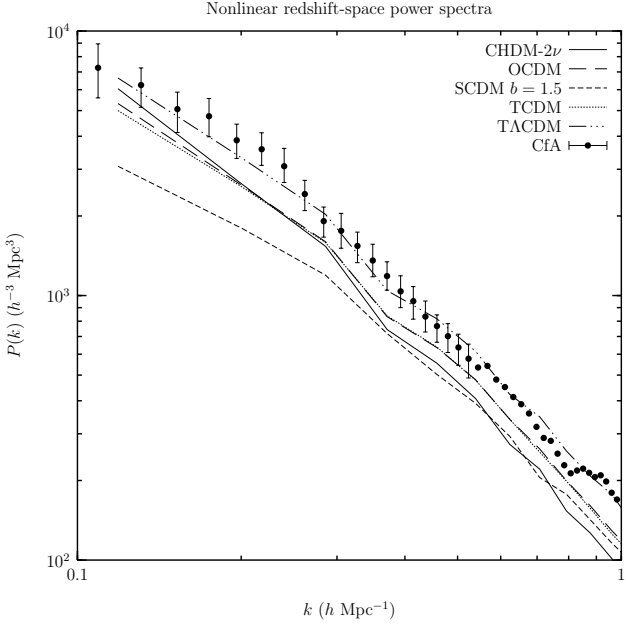

\ifref\else
\begin{fig}{power_rs}
\fi
  \caption[Redshift space dark matter power spectrum]
          {Redshift space power spectrum, compared to the combined CfA2 and
           SSRS2 redshift space power spectrum \protect\cite{dacosta:1994b}.
           Notice that, while $\Lambda$CDM is a good match to this power
           spectrum, there is no room for galaxy formation or velocity bias.}
  \label{fig:rspow}
\ifref\else
\end{fig}
\fi
\end{figure}

The APM power spectrum is not the only power spectrum that has been measured.
However, to compare to other measurements, it is usually required to calculate
model redshift space power spectra.  Going to redshift space significantly
reduces power on scales of interest, since typical dispersion velocities of
1000--2000 $\kms$ in clusters correspond to a scatter of 10--20 $\hmpc$ in
distance.  In performing this operation on particles, the power should be
viewed as a lower bound, because there may be significant velocity bias
(\citeNP{carlberg:1990}; \citeNP{summers:1995}), meaning the power perhaps
shouldn't be supressed \emph{quite} as much, and galaxy formation will further
raise the power.  Figure \ref{fig:rspow}%
shows the models' redshift space power spectra, compared to
the combined CfA2 and SSRS2 redshift space power spectrum \cite{dacosta:1994b}.
Given our choices of model normalization and cosmological parameters, the
T$\Lambda$CDM matter power spectrum is nicely consistent with the observed
galaxy power spectrum, but that leaves \emph{no} room for galaxy formation or
velocity bias effects.  As for the real-space nonlinear power spectrum
comparison (figure~\ref{fig:nlpow}), this requires significant antibiasing for
T$\Lambda$CDM on scales of 0.3--1 $\hmpcinv$.  Note that undersampling the
velocity field will miss the large velocities by making the halos physically
larger, so it does not make sense to perform redshift space comparisons on the
large volume suite.

The simplest halo-related quantity to investigate is the number density of
bound objects as a function of mass.  Such `mass functions' and close relatives
such as the X-ray temperature function (as in \citeNP{eke:1996}, for example)
are often estimated from the Press-Schechter approximation instead of from
simulations.  Though it has been checked against scale-free simulations
\cite{efstathiou:1988a,bond:1991,lacey:1994} and
against specific SCDM, $\Lambda$CDM and CHDM models
\cite{carlberg:1989,jain:1994,klypin:1995,walter:1996,bond:1996b}, previous
studies have focused only on a narrow range of masses, typically at the cluster
scale.  With our large simulations, we can check the approximation over four
orders of magnitude in mass.
The Press-Schechter formula we use is \citeN{klypin:1995},
equations (1--2), evaluated at $z=0$:
\begin{equation}
\label{eqn:press1}
N(>M)=\sqrt{\frac{2}{\upi}}\frac{\delc}{\alpha_{\mathrm{m}}}%
       \int_r^\infty\frac{\epsilon(r')}{\sigma^3(r')}%
           \exp\left[\frac{-\delc^2}{2\sigma^2(r')}\right]%
                 \frac{{\mathrm{d}}r'}{{r'}^3},
\end{equation}
where
\begin{equation}
\label{eqn:press2}
\epsilon(r)=\frac{1}{2\upi}%
         \int_0^\infty k^3P(k)W(kr)\frac{{\mathrm{d}}W(kr)}{{\mathrm{d}}(kr)}%
            {\,\mathrm{d}}k,
\end{equation}
\begin{equation}
\label{eqn:press3}
\sigma^2(r)=\frac{1}{2\upi^2}\int_0^\infty k^2P(k)W^2(kr){\,\mathrm{d}}k,
\end{equation}
\begin{equation}
\label{eqn:press4}
W(x)=\left\{
  \begin{array}{ll}
    \frac{3}{x^3}\left[\sin(x) - x\cos(x)\right]&\mbox{tophat}\\
    e^{-x^2/2}                                  &\mbox{Gaussian}
  \end{array}\right.,
\end{equation}
\begin{equation}
\label{eqn:press5}
r = \left(\frac{M} {\alpha_{\mathrm{m}}\rhocrit\Omega_0}\right)^{\frac{1}{3}}
\end{equation}
and $\alpha_{\mathrm{m}}$ is $4\upi/3$ for a tophat window function, and
$(2\upi)^{3/2}$ for a Gaussian window function.

\begin{figure*}
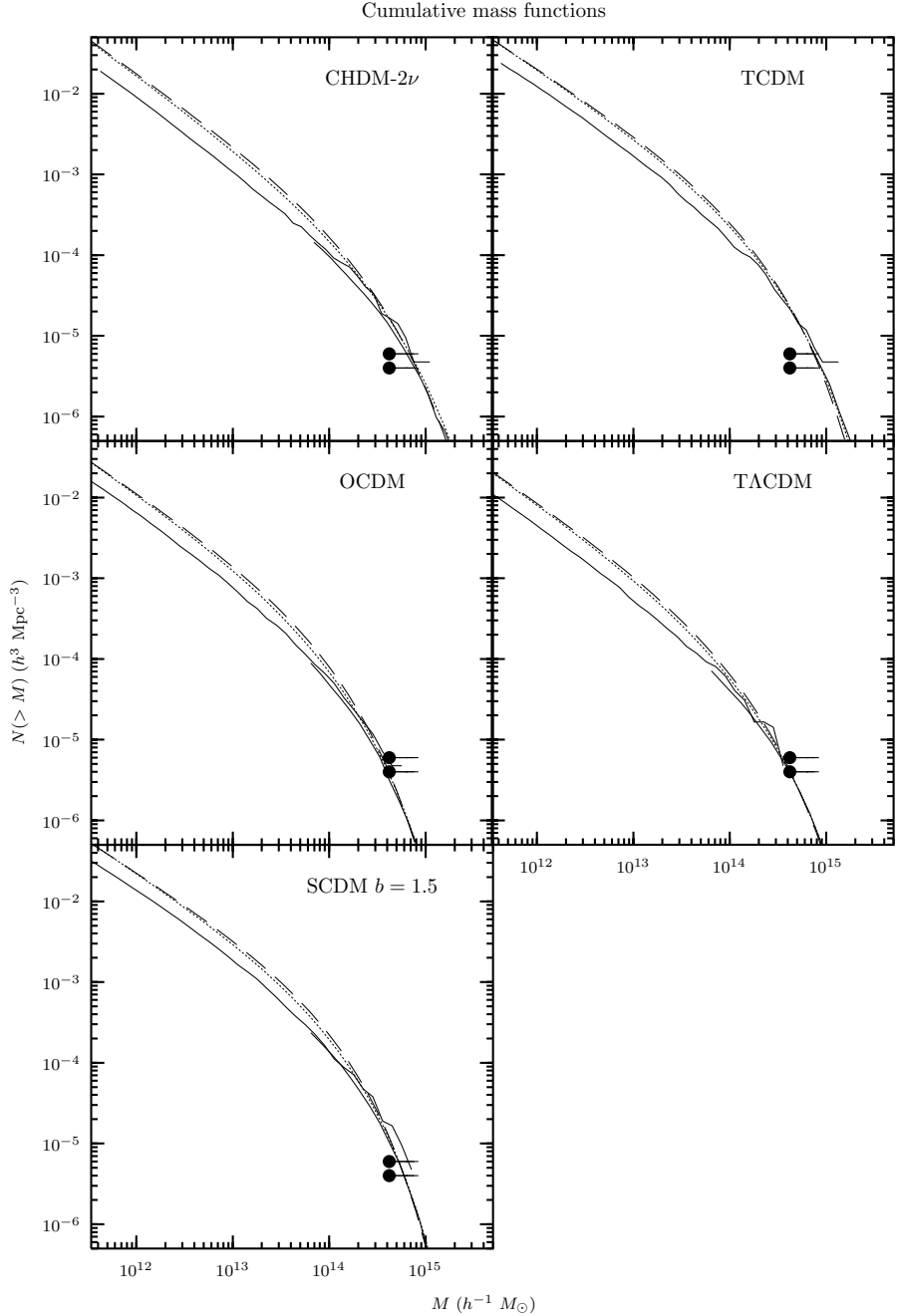

\ifref\else
\begin{figlong}{mf}
\fi
  \caption[Cumulative halo mass functions, with Press-Schechter fits]
          {Cumulative halo mass functions, with Press-Schechter fits.  In each
           panel, the relevant mass functions estimated from the two simulation
           suites is shown by the full curves.  Small mass cuts have been
           applied at $M=3.4\times 10^{11}$ $\hmsun$ and
           $2.2\times 10^{13}$ $\hmsun$, for the large and small
           volume simulations, respectively.  Each panel also shows
           Gaussian (dashed curves) and tophat (dotted curves) Press-Schechter
           mass functions,
           with $\delct$ and $\delcg$ adjusted
           to agree with the large-volume simulations
           at $5.5\times 10^{14}$ $\hmsun$.  The values of $\delct$ and $\delcg$
           used are given in table~\ref{tbl:psvirnorm}.  The data points
           correspond to the observations of
           BGGMM93\protect\nocite{biviano:1993}
           and WEF93\protect\nocite{white:1993}, as in
           figure~\protect\ref{fig:ps}.}
  \label{fig:mf}
\ifref\else
\end{figlong}
\fi
\end{figure*}
Figure~\ref{fig:mf} shows the cumulative mass functions estimated from both
suites of simulations, and
\begin{table}
  \tblcaptraw{
    \begin{tabular}{lcc}
    Model & $\delct$ & $\delcg$ \\
    \hline
    CHDM-2$\nu$   & 1.571 & 1.273 \\
    OCDM          & 1.693 & 1.293 \\
    SCDM $b=1.5$  & 1.672 & 1.236 \\
    TCDM          & 1.630 & 1.252 \\
    T$\Lambda$CDM & 1.732 & 1.355 \\
    \hline
    \end{tabular}
  }
  {Press-Schechter fits to simulation mass functions}
  {tbl:psvirnorm}
  {\def\baselinestretch{1.0} \Huge\small 
    {$\delct$ and $\delcg$ have been chosen to get the same number density of
    clusters with $M>5.5\times 10^{14}$ $\hmsun$ as the large-volume
    simulation, for each model.}
  }
\end{table}
table~\ref{tbl:psvirnorm} shows the Press-Schechter
parameters used in that figure.  Note that in the overlapping region, the two
sets of simulation mass functions are consistent, and that \emph{the high
resolution results are a significant factor of 1.5--2 below the Press-Schechter
estimates for all models at the intermediate mass of $10^{13}$ $\hmsun$ and
below.}

This result has been verified recently by other groups.  \citeN{bryan:1998}
see a somewhat stronger discrepancy at $10^{14}$ $\hmsun$, using a spherical
overdensity method, for cosmological parameters very close to our CHDM-2$\nu$,
OCDM and SCDM choices (though with substantially larger grid cell sizes for
OCDM and SCDM).  \citeN{somerville:1998a} also see an equivalent discrepancy
in the differential multiplicity function 
$n(M)$ at $z=0$ in the $\tau$CDM cosmology, for which the power spectrum is
very similar to our CHDM-2$\nu$.  This result depends upon a completely
independent simulation \cite{jenkins:1997} modeled using adaptive P$^3$M, with
halos identified using the Friends-of-Friends method.  The halo mass function
for the SCDM model from the \citeN{jenkins:1997} simulations is virtually
identical to ours (G. Lemson, private communication).
Given these confirmations, we do not believe that the medium-mass
discrepancy we see is an artifact of our simulation method or
halo finding algorithm.

\begin{figure*}
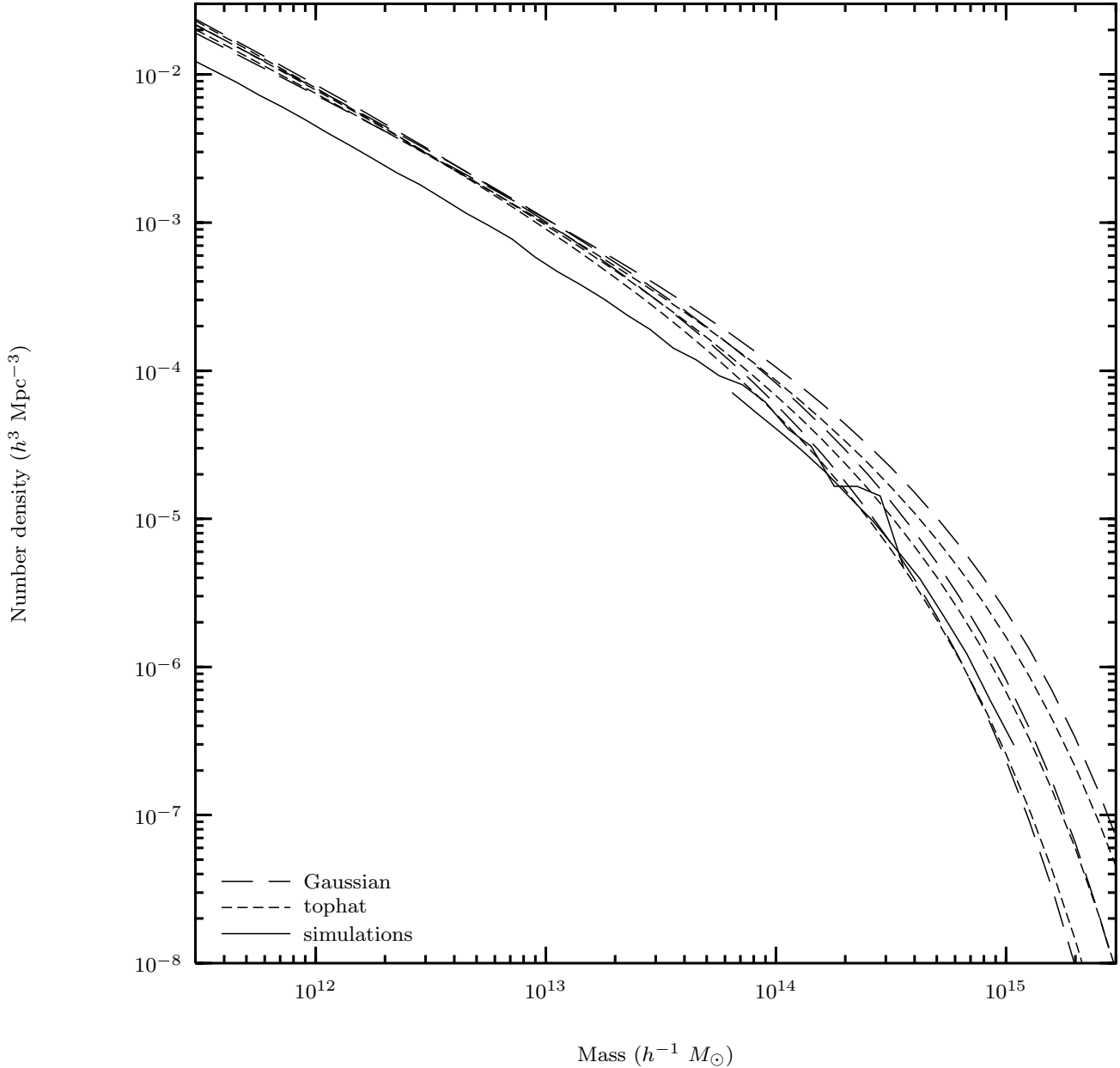

\ifref\else
\begin{figwide}{delc}
\fi
  \caption[Press-Schechter mass functions for T$\Lambda$CDM]
          {Press-Schechter mass functions for T$\Lambda$CDM.  The high and
           low density T$\Lambda$CDM mass functions from simulations are
           shown in the solid curves.  From top to bottom at $M=10^{15}$
           $\hmsun$, the dashed curves show Press-Schechter mass
           functions with Gaussian filters for $\protect\delcg=1.0$, $1.2$,
           and $1.4$, and the dotted curves show tophat filters for
           $\protect\delct=1.4$, $1.6$, and $1.8$.  Press-Schechter mass
           functions can be made to agree with our simulations for masses above
           about $5\times 10^{13}$ $\hmsun$, but not for masses
           smaller than that.}
  \label{fig:delc}
\ifref\else
\end{figwide}
\fi
\end{figure*}

As figure~\ref{fig:delc} shows, the intermediate and
low mass discrepancy \emph{cannot} be fixed by adjusting the value of $\delc$,
particularly at a mass of $~\sim 5\times 10^{12}$ $\hmsun$, where
the curves cross.  Our values of $\delct$ and $\delcg$ are consistent with
\citeN{borgani:1997a},
except we find that the CHDM-2$\nu$ $\delc$'s are not significantly different
from the other models.  The $\delct$ values we find for the
tophat case are consistent with the spherical collapse model.

The simulation mass functions in figure~\ref{fig:mf} fall below the
Press-Schechter predictions for most of their range.  For example, in the
SCDM high-resolution run, only 60 per cent of the particles are within halos
with $M>3.4\times10^{11}$ $\hmsun$ at $z=0$.  The Press-Schechter prediction
is only very slightly larger, about 62 per cent for $\delct=1.672$ and 65 per
cent for the spherical collapse value $\delct=1.686$.  The mass deficit due to
smaller abundance of low-mass halos halos in the simulations is almost
completely compensated for by a small excess of very large clusters.  The
Press-Schechter approximation assumes that all the mass must be in halos of
\emph{some} size, and this analysis indicates that a significant fraction of
the mass of the universe should be in small halos.  The Press-Schechter
approximation indicates that for SCDM, as much as 20 per cent of the mass is
in halos as small as $10^9$ $\hmsun$, which is not identifiable in any
present-day cosmological simulation.  The simulations show a significant amount
of matter that is not in collapsed objects.  Most of the mass lies in filaments
connecting the clusters, many of which have only a few identified halos on
them.  It is conceivable that much of this mass may be unresolved halos, since
any $N$-body simulation must have a resolution and/or timestep limit below
which forces are `soft,' resulting in disruption of structure smaller than the
limit.  Such a mechanism must be present, since filament halos are necessarily
not very big, but it is not clear how much of the mass that can account for.

Our two low-$\Omega_0$ models produce fewer clusters in simulations than the
other models (figure~\ref{fig:mf}).  If X-ray temperature cluster masses are
correct, this presents no problem for those models.  However, if
the indications of larger cluster masses from gravitational lensing
are correct, the low-$\Omega_0$ models require revision by using less tilt
(in the case of T$\Lambda$CDM) or a larger value of $H_0$. The former 
would help lessen the disagreement with high-multipole cosmic microwave
background measurements (figure~\ref{fig:cmb}), as a weaker tilt would raise
the first Doppler peak, but will lead to the need for even stronger anti-bias
to reconcile small-scale power with the APM observations. If X-ray temperature
masses are correct, our parameters for TCDM and CHDM-2$\nu$ produce too many
clusters. For CHDM-2$\nu$, the normalization used here was actually about 10
per cent higher than the preferred four-year \cobe\ normalization, so reducing
the normalization by this factor would probably be enough, though this will
exacerbate early structure formation problems.  For TCDM, the only options are
to either increase the tilt, which is highly disfavored by the small-angle
cosmic microwave background data as already noted, or further reduce the Hubble
parameter, which is also strongly disfavored by observations.

The statements above all take the \cobe\ normalization as a fixed constraint.
Alternatively, we could turn the problem around and use the clusters to
determine normalizations and tilts, with $H_0$ (and $\Omega_0$ and $\oml$) as a
given.  This is explored further in
\ifpaper
  \citeN{gross:clusters}.
\else
  chapter~\ref{chap:clusters}.
\fi

One would now like to investigate statistics such as correlation functions,
void probability functions \cite{ghigna:1997}, shape statistics
\cite{dave:1997}, and other sophisticated statistics.  However, to compare to
observations, we need to know how many galaxies form in each halo.  Previous
studies (KNP97\nocite{KNP:1997}; \citeNP{nolthenius:1997,ghigna:1997}, for
example) have used ad hoc `breakup' prescriptions to assign galaxies to halos. 
We intend to populate our halos with galaxies using a more physically motivated
approach (as in \citeNP{kauffmann:1997}) based on semi-analytic models including
simplified treatments of gas processes, star formation, supernova feedback, and
galaxy-galaxy merging \cite{somerville:thesis,somerville:1998b}. 
As a result, we do not attempt to include any complicated galaxy identification
algorithms here.

For certain statistics, one can partially compensate for the effect of
overmerging by \emph{mass weighting}.  This 
approach is less than ideal because it does not restore the small-scale
\emph{spatial} information lost in the overmerging process.  Mass weighting is
equivalent to presuming a halo contains a number of galaxies proportional to
its mass, and putting all the galaxies at the center of the halo.  In effect,
this clears out regions of space around the largest halos' centers, equal to
their radii, and therefore loses information on scales smaller than the largest
halo radius (typically 2--3 $\hmpc$).  Since very massive halos are rare
objects for physically interesting cosmological models, all mass weighted
statistics must be unduly influenced by small-number statistical noise.

\begin{figure}
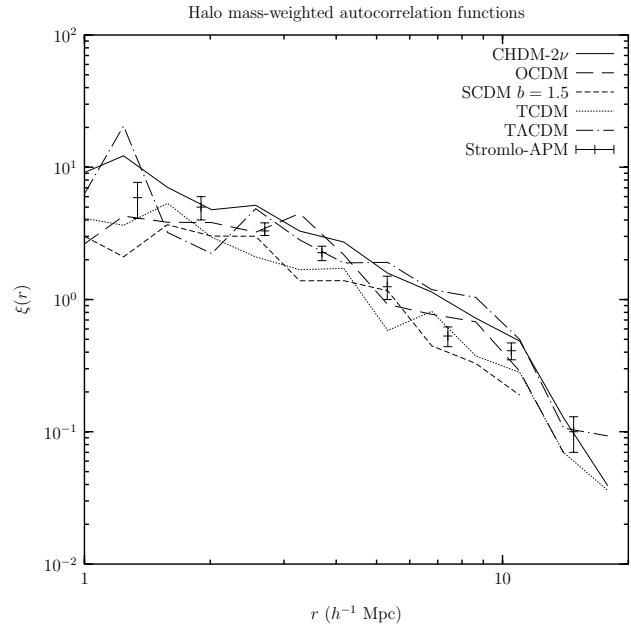

\ifref\else
\begin{fig}{cf}
\fi
  \caption[Halo mass-weighted correlation function]
          {Halo mass-weighted correlation function.  Only halos with
           mass above $3\times 10^{11}$ $\hmsun$ are included, and
           all pairs are weighted by the product of the two masses.  Such
           a weighting is a simple countermeasure for the overmerging problem.
           The error bars are the Stromlo-APM autocorrelation function
           \cite{loveday:1995}.}
  \label{fig:cf}
\ifref\else
\end{fig}
\fi
\end{figure}

We calculate the mass-weighted autocorrelation function for the high
resolution runs, and the results are shown in figure~\ref{fig:cf}.
In this figure, a halo mass cut of $M=3\times 10^{11}$ $\hmsun$ was
used, although the mass weighting makes it insensitive to the mass cut.  The
mass weighting creates a spread in the correlation values large enough to
prevent the test from discriminating among models.  To within the spread
visible in figure~\ref{fig:cf}, all models are roughly consistent with the
Stromlo-APM autocorrelation function \cite{loveday:1995}. However, there are a
few trends visible in the figure.  SCDM and TCDM are systematically lower in
amplitude than the other models, but the effect is not very significant given
the spread.  

\section{Conclusions}
\label{sec:conclusion}

We have run two suites of simulations with 57 million cold particles in
boxes of 75 and 300 $\hmpc$, with the goal of studying interesting
variants of the CDM family of cosmological models.
In this paper, we have made preliminary comparisons of the $z=0$
simulation outputs to data for all models.
In \citeN{smith:1998}, we used the lower resolution suite, plus some additional
simulations, to generalize the Peacock \& Dodds
\citeyear{peacock:1994,peacock:1996} procedure for recovery of the linear
power spectrum corresponding to a given cosmological model from observational
data.  In \citeN{wechsler:1998}, we showed that the most massive halos at
redshifts $z\sim 3$, or objects that trace their distribution, can account for
the observed clustering of Lyman-break objects \cite{steidel:1998} for all
cosmologies except SCDM.  More detailed comparisons with
observations require assumptions about galaxy formation and will be treated
in subsequent work.  Subject to the usual caveats about the
uncertainty of galaxy formation, we reach the following
conclusions in the present paper:
\begin{enumerate}
\item Based on the results of KPH96\nocite{klypin:1996a}, who found that
      $\Lambda$CDM models with $\Omega_0\sim0.3$ would require strong scale
      dependent anti-bias in order to be consistent with the APM power spectrum
      \cite{baugh:1994}, we investigated a variant of the $\Lambda$CDM
      model with $\Omega_0=0.4$ and a tilt of $n=0.9$. We find that this
      model still requires large antibias of $b^2\equiv
      P_{\mathrm{APM gal}}/P_{\mathrm{dm}} \sim 0.5$ at $k=1$ $\hmpcinv$.
      Even in a simple model in which galaxies are extremely anti-biased with
      respect to dark matter halos, the problem persists on scales of $r\sim
      6$$\hmpc$, because this scale is larger than the size of individual
      halos. To get anti-bias on these scales, there would have to be many
      `barren' halos containing \emph{no} galaxies.  OCDM and TCDM are
      only slightly better, still requiring a strong antibias of $b^2\sim 0.6$.  Other models considered
      require a weaker antibias at that scale.  
\item The T$\Lambda$CDM dark matter redshift space power spectrum agrees very
      well with the redshift space galaxy power spectrum from CfA2+SSRS2
      \cite{dacosta:1994b}.  This leaves no room for the `positive' bias
      expected in normal
      galaxy formation, or for velocity biases.  For comparison, OCDM and TCDM
      each have room for a modest bias of $b^2\sim 1.2$ at $k=0.5$ $\hmpcinv$,
      and CHDM-2$\nu$ and SCDM each need $b^2\sim 1.5$.
\item All models considered here are consistent with the Stromlo-APM real-space
      correlation function \cite{loveday:1995} on scales of 2--20 $\hmpc$,
      largely due to a large spread in the model estimates of the
      correlation function because of mass weighting and small-number
      statistics for large mass objects.
\item The Press-Schechter approximation fits the abundance of cluster-mass
      halos very well, with top-hat $\delct$=1.57--1.73 and Gaussian
      $\delcg=$1.27--1.35.  However, it overpredicts the number density of
      galaxy and small group mass objects by a factor of $\sim 2$, only weakly
      dependent on cosmology, and very weakly dependent on $\delc$.  On mass
      scales of $\sim 5\times 10^{12}$ $\hmsun$, it is not possible to
      compensate for the discrepancy by adjusting $\delc$ within reasonable
      bounds.
\end{enumerate}

In summary, we conclude that none of the models we have investigated can be
strongly ruled out by the kind of analysis performed here. The CHDM-2$\nu$
model gives the best overall agreement with the linear and non-linear tests we
have considered here, assuming that galaxies are positively biased with respect
to the dark matter.
\citeN{gawiser:1998} have shown that a similar CHDM model with $\omh=0.2$ in
$N_\nu=1$ neutrino species is a much better fit to microwave background
and galaxy distribution data than any other popular cosmological model.
Preliminary analysis based on the dark matter alone has shown that
the related CHDM-2$\nu$ model considered in this paper is plausibly consistent
with high redshift observations
of Lyman-break galaxies \cite{wechsler:1998} and damped Lyman-$\alpha$ systems
\cite{klypin:1994}, but it remains to be seen whether this model will produce
enough early galaxy formation once a more realistic treatment of gas processes
and star formation is included. More detailed modelling of galaxy formation
will also be necessary to determine whether the small-scale clustering
properties of the low-$\Omega_0$ models are indeed inconsistent with the
observations. In any case we conclude that models with $\Omega_0\sim0.5$ are in
better overall agreement with the observations than the lower values
($\Omega_0\sim0.2$--$0.3$) usually considered (e.g. \citeNP{jenkins:1997}). A
powerful constraint on $\Omega_0$, the evolution of cluster abundance with
redshift, will be considered in a companion paper \cite{gross:clusters}.

\section{Acknowledgements}
MAKG and JRP were supported by NSF and NASA grants at UCSC, RSS was supported
by a GAANN fellowship, and JH and AK were supported by NSF and NASA grants
at NMSU.  The simulations were run on the IBM SP2 at the Cornell Theory Center,
Cornell University, Ithaca, NY, USA and computer time for analysis was provided
by Sandra Faber and the DEIMOS project at UCSC.


\ifpaper
  \bibliographystyle{mnras}
  \bibliography{mnrasmnemonic,thesis}

\begin{thebibliography}{}

\bibitem[\protect\citeauthoryear{Bardeen et~al.}{Bardeen
  et~al.}{1986}]{BBKS:1986}
Bardeen J.~M., Bond J.~R., Kaiser N.,  Szalay A.~S., 1986, ApJ, 304, 15 (BBKS)

\bibitem[\protect\citeauthoryear{Bartlett et~al.}{Bartlett
  et~al.}{1995}]{bartlett:1995}
Bartlett J.~G., Blanchard A., Silk J.,  Turner M.~S., 1995, Science, 267, 980

\bibitem[\protect\citeauthoryear{Baugh \& Efstathiou}{Baugh \&
  Efstathiou}{1994}]{baugh:1994}
Baugh C.~M.,  Efstathiou G., 1994, MNRAS, 267, 323

\bibitem[\protect\citeauthoryear{Bernardeau et~al.}{Bernardeau
  et~al.}{1995}]{bernardeau:1995}
Bernardeau F., Juszkiewicz R., Dekel A.,  Bouchet F., 1995, MNRAS, 274, 20

\bibitem[\protect\citeauthoryear{Biviano et~al.}{Biviano
  et~al.}{1993}]{biviano:1993}
Biviano A., Girardi M., Giuricin G., Mardirossian F.,  Mezzetti M., 1993, ApJ,
  411, L13 (BGGMM93)

\bibitem[\protect\citeauthoryear{Blumenthal et~al.}{Blumenthal
  et~al.}{1984}]{blumenthal:1984}
Blumenthal G.~R., Faber S.~M., Primack J.~R.,  Rees M.~J., 1984, Nat, 517

\bibitem[\protect\citeauthoryear{Bond et~al.}{Bond et~al.}{1991}]{bond:1991}
Bond J.~R., Cole S., Efstathiou G.,  Kaiser N., 1991, ApJ, 379, 440

\bibitem[\protect\citeauthoryear{Bond \& Myers}{Bond \&
  Myers}{1996}]{bond:1996b}
Bond J.~R.,  Myers S.~T., 1996, ApJS, 103, 41

\bibitem[\protect\citeauthoryear{Borgani et~al.}{Borgani
  et~al.}{1997a}]{borgani:1997b}
Borgani S., Gardini A., Girardi M.,  Gottl\"ober S., 1997a, New Astronomy, 2,
  119

\bibitem[\protect\citeauthoryear{Borgani et~al.}{Borgani
  et~al.}{1997b}]{borgani:1997a}
Borgani S. et~al., 1997b, New Astronomy, 1, 299

\bibitem[\protect\citeauthoryear{Bryan \& Norman}{Bryan \&
  Norman}{1998}]{bryan:1998}
Bryan G.~L.,  Norman M.~L., 1998, ApJ, 495, 80

\bibitem[\protect\citeauthoryear{Bunn \& White}{Bunn \&
  White}{1997}]{bunn:1997}
Bunn E.~F.,  White M., 1997, ApJ, 480, 6

\bibitem[\protect\citeauthoryear{Burles \& Tytler}{Burles \&
  Tytler}{1997a}]{burles:1997a}
Burles S.,  Tytler D., 1997a, apj in press, astro-ph/9712108

\bibitem[\protect\citeauthoryear{Burles \& Tytler}{Burles \&
  Tytler}{1997b}]{burles:1997b}
Burles S.,  Tytler D., 1997b, apj in press, astro-ph/9712109

\bibitem[\protect\citeauthoryear{Carlberg \& Couchman}{Carlberg \&
  Couchman}{1989}]{carlberg:1989}
Carlberg R.~G.,  Couchman H.~M.~P., 1989, ApJ, 340, 47

\bibitem[\protect\citeauthoryear{Carlberg, Couchman, \& Thomas}{Carlberg
  et~al.}{1990}]{carlberg:1990}
Carlberg R.~G., Couchman H.~M.~P.,  Thomas P.~A., 1990, ApJ, 352, L29

\bibitem[\protect\citeauthoryear{Chaboyer et~al.}{Chaboyer
  et~al.}{1997}]{chaboyer:1997}
Chaboyer B., Demarque P., Kernan P.~J.,  Krauss L.~M., 1997, preprint,
  astro-ph/9706128

\bibitem[\protect\citeauthoryear{Couchman}{Couchman}{1991}]{couchman:1991}
Couchman H.~M.~P., 1991, ApJ, 368, L23

\bibitem[\protect\citeauthoryear{da~Costa et~al.}{da~Costa
  et~al.}{1994}]{dacosta:1994b}
da~Costa L.~N., Vogeley M.~S., Geller M.~J., Huchra J.~P.,  Park C., 1994, ApJ,
  437, L1

\bibitem[\protect\citeauthoryear{Dav\'e et~al.}{Dav\'e
  et~al.}{1997}]{dave:1997}
Dav\'e R., Hellinger D., Primack J.~R., Nolthenius R.,  Klypin A., 1997, MNRAS,
  284, 607

\bibitem[\protect\citeauthoryear{Davis et~al.}{Davis et~al.}{1985}]{DEFW:1985}
Davis M., Efstathiou G., Frenk C.~S.,  White S.~D.~M., 1985, ApJ, 292, 371

\bibitem[\protect\citeauthoryear{Dekel et~al.}{Dekel et~al.}{1997}]{dekel:1997}
Dekel A. et~al., 1997, in preparation

\bibitem[\protect\citeauthoryear{Dekel \& Rees}{Dekel \&
  Rees}{1994}]{dekel:1994b}
Dekel A.,  Rees M.~J., 1994, ApJ, 422, L1

\bibitem[\protect\citeauthoryear{Efstathiou et~al.}{Efstathiou
  et~al.}{1988}]{efstathiou:1988a}
Efstathiou G., Frenk C.~S., White S.~D.~M.,  Davis M., 1988, MNRAS, 235, 715

\bibitem[\protect\citeauthoryear{Eke, Cole, \& Frenk}{Eke
  et~al.}{1996}]{eke:1996}
Eke V.~R., Cole S.,  Frenk C.~S., 1996, MNRAS, 282, 263

\bibitem[\protect\citeauthoryear{Frenk et~al.}{Frenk et~al.}{1998}]{frenk:1998}
Frenk C.~S. et~al., 1998, submitted to ApJ

\bibitem[\protect\citeauthoryear{Gawiser \& Silk}{Gawiser \&
  Silk}{1998}]{gawiser:1998}
Gawiser E.,  Silk J., 1998, Science, in press

\bibitem[\protect\citeauthoryear{Gelb \& Bertschinger}{Gelb \&
  Bertschinger}{1994}]{gelb:1994a}
Gelb J.~M.,  Bertschinger E., 1994, ApJ, 436, 467

\bibitem[\protect\citeauthoryear{Ghigna et~al.}{Ghigna
  et~al.}{1997}]{ghigna:1997}
Ghigna S., Borgani S., Tucci M., Bonometto S.~A., Klypin A.,  Primack J.~R.,
  1997, ApJ, 479, 580

\bibitem[\protect\citeauthoryear{G\'orski et~al.}{G\'orski
  et~al.}{1996}]{gorski:1996a}
G\'orski K.~M., Banday A.~J., Bennett C.~L., Hinshaw G., Kogut A., Smoot G.~F.,
   Wright E.~L., 1996, ApJ, 464, L11

\bibitem[\protect\citeauthoryear{Gratton et~al.}{Gratton
  et~al.}{1997}]{gratton:1997}
Gratton R.~G., Pecci F.~F., Carretta E., Clementini G., Corsi C.~E.,  Lattantzi
  M., 1997, preprint, astro-ph/9704150

\bibitem[\protect\citeauthoryear{Gross}{Gross}{1997}]{gross:thesis}
Gross M.~A.~K., 1997, Ph.D. thesis, University of California, Santa Cruz,
  http://fozzie.gsfc.nasa.gov/\#dissertation

\bibitem[\protect\citeauthoryear{Gross et~al.}{Gross
  et~al.}{1998}]{gross:clusters}
Gross M.~A.~K., Somerville R.~S., Primack J.~R., Borgani S.,  Girardi M., 1998,
  in preparation

\bibitem[\protect\citeauthoryear{Henry \& Arnaud}{Henry \&
  Arnaud}{1991}]{henry:1991}
Henry J.~P.,  Arnaud K.~A., 1991, ApJ, 372, 410

\bibitem[\protect\citeauthoryear{Hockney \& Eastwood}{Hockney \&
  Eastwood}{1988}]{hockney:1988}
Hockney R.~W.,  Eastwood J.~W., 1988, Computer Simulation Using Particles.
\newblock Institute of Physics Publishing, Philadelphia

\bibitem[\protect\citeauthoryear{Holtzman}{Holtzman}{1989}]{holtzman:1989}
Holtzman J.~A., 1989, ApJS, 71, 1

\bibitem[\protect\citeauthoryear{Hu \& Sugiyama}{Hu \&
  Sugiyama}{1996}]{hu:1996}
Hu~W.,  Sugiyama N., 1996, ApJ, 471, 542

\bibitem[\protect\citeauthoryear{Hu \& White}{Hu \& White}{1997}]{hu:1997b}
Hu~W.,  White M., 1997, ApJ, 486, L1

\bibitem[\protect\citeauthoryear{Jain \& Bertschinger}{Jain \&
  Bertschinger}{1994}]{jain:1994}
Jain B.,  Bertschinger E., 1994, ApJ, 431, 495

\bibitem[\protect\citeauthoryear{Jenkins et~al.}{Jenkins
  et~al.}{1997}]{jenkins:1997}
Jenkins A. et~al., 1997, ApJ, in press, astro-ph/9709010 v2

\bibitem[\protect\citeauthoryear{Kates, Kotok, \& Klypin}{Kates
  et~al.}{1991}]{kates:1991}
Kates R.~E., Kotok E.~V.,  Klypin A., 1991, A\&A, 243, 295

\bibitem[\protect\citeauthoryear{Kauffmann \& Charlot}{Kauffmann \&
  Charlot}{1994}]{kauffmann:1994}
Kauffmann G.,  Charlot S., 1994, ApJ, 430, L97

\bibitem[\protect\citeauthoryear{Kauffmann, Nusser, \& Steinmetz}{Kauffmann
  et~al.}{1997}]{kauffmann:1997}
Kauffmann G., Nusser A.,  Steinmetz M., 1997, MNRAS, 286, 795

\bibitem[\protect\citeauthoryear{Klypin et~al.}{Klypin
  et~al.}{1995a}]{klypin:1994}
Klypin A., Borgani S., Holtzman J.,  Primack J., 1995a, ApJ, 444, 1

\bibitem[\protect\citeauthoryear{Klypin et~al.}{Klypin
  et~al.}{1995b}]{klypin:1995}
Klypin A., Borgani S., Holtzman J.,  Primack J., 1995b, ApJ, 444, 1

\bibitem[\protect\citeauthoryear{Klypin, Gottl\"ober, \& Kravtsov}{Klypin
  et~al.}{1997}]{klypin:1997}
Klypin A., Gottl\"ober S.,  Kravtsov A.~V., 1997, preprint, astro-ph/9708191

\bibitem[\protect\citeauthoryear{Klypin et~al.}{Klypin
  et~al.}{1993}]{KHPR:1993}
Klypin A., Holtzman J., Primack J.,  Reg\"{o}s E., 1993, ApJ, 416, 1

\bibitem[\protect\citeauthoryear{Klypin, Nolthenius, \& Primack}{Klypin
  et~al.}{1997}]{KNP:1997}
Klypin A., Nolthenius R.,  Primack J., 1997, ApJ, 474, 533 (KNP97)

\bibitem[\protect\citeauthoryear{Klypin, Primack, \& Holtzman}{Klypin
  et~al.}{1996}]{klypin:1996a}
Klypin A., Primack J.,  Holtzman J., 1996, ApJ, 466, 1 (KPH96)

\bibitem[\protect\citeauthoryear{Kolatt \& Dekel}{Kolatt \&
  Dekel}{1997}]{kolatt:1997}
Kolatt T.,  Dekel A., 1997, ApJ, 479, 592

\bibitem[\protect\citeauthoryear{Kravtsov, Klypin, \& Khokhlov}{Kravtsov
  et~al.}{1997}]{kravtsov:1997}
Kravtsov A.~V., Klypin A.,  Khokhlov A.~M., 1997, ApJS, 111, 73

\bibitem[\protect\citeauthoryear{Lacey \& Cole}{Lacey \&
  Cole}{1994}]{lacey:1994}
Lacey C.,  Cole S., 1994, MNRAS, 271, 676

\bibitem[\protect\citeauthoryear{Liddle et~al.}{Liddle
  et~al.}{1996}]{liddle:1996c}
Liddle A.~R., Lyth D.~H., Roberts D.,  Viana P.~T.~P., 1996, MNRAS, 278, 644
  (LLRV96)

\bibitem[\protect\citeauthoryear{Lineweaver \& Barbosa}{Lineweaver \&
  Barbosa}{1997}]{lineweaver:1997b}
Lineweaver C.~H.,  Barbosa D., 1997, preprint, astro-ph/9706077

\bibitem[\protect\citeauthoryear{Loveday et~al.}{Loveday
  et~al.}{1995}]{loveday:1995}
Loveday J., Maddox S.~J., Efstathiou G.,  Petersen B.~A., 1995, ApJ, 442, 457

\bibitem[\protect\citeauthoryear{Miralda-Escud\'e \& Babul}{Miralda-Escud\'e \&
  Babul}{1995}]{miralda-escude:1995}
Miralda-Escud\'e J.,  Babul A., 1995, ApJ, 449, 18

\bibitem[\protect\citeauthoryear{Monaco}{Monaco}{1995}]{monaco:1995}
Monaco P., 1995, ApJ, 447, 23

\bibitem[\protect\citeauthoryear{Navarro, Frenk, \& White}{Navarro
  et~al.}{1996}]{navarro:1996b}
Navarro J.~F., Frenk C.~S.,  White S.~D.~M., 1996, preprint, astro-ph/9611107
  v2

\bibitem[\protect\citeauthoryear{Netterfield et~al.}{Netterfield
  et~al.}{1997}]{netterfield:1997}
Netterfield C.~B., Devlin N., Jarosik L., Page L.,  Wollack E.~J., 1997, ApJ,
  474, 47

\bibitem[\protect\citeauthoryear{Nolthenius, Klypin, \& Primack}{Nolthenius
  et~al.}{1997}]{nolthenius:1997}
Nolthenius R., Klypin A.,  Primack J.~R., 1997, ApJ, 480, 43

\bibitem[\protect\citeauthoryear{Nusser \& Dekel}{Nusser \&
  Dekel}{1993}]{nusser:1993}
Nusser A.,  Dekel A., 1993, ApJ, 405, 437

\bibitem[\protect\citeauthoryear{Peacock \& Dodds}{Peacock \&
  Dodds}{1994}]{peacock:1994}
Peacock J.~A.,  Dodds S.~J., 1994, MNRAS, 267, 1020 (PD94)

\bibitem[\protect\citeauthoryear{Peacock \& Dodds}{Peacock \&
  Dodds}{1996}]{peacock:1996}
Peacock J.~A.,  Dodds S.~J., 1996, MNRAS, 280, L19 (PD96)

\bibitem[\protect\citeauthoryear{Pen}{Pen}{1996}]{pen:1996}
Pen U.~L., 1996, astro-ph/9610147

\bibitem[\protect\citeauthoryear{Platt et~al.}{Platt et~al.}{1997}]{platt:1997}
Platt S.~R., Kovac J., Dragovan M., Petersen J.~B.,  Ruhl J.~E., 1997, ApJ,
  475, L1

\bibitem[\protect\citeauthoryear{Pogosyan \& Starobinsky}{Pogosyan \&
  Starobinsky}{1995}]{pogosyan:1995a}
Pogosyan D.~Y.,  Starobinsky A.~A., 1995, in Mucket J.~P., Gottl\"ober S.,
  M\"uller V., ed, International Workshop on Large Scale Structure in the
  Universe.
\newblock World Scientific, River Edge, New Jersey

\bibitem[\protect\citeauthoryear{Press \& Schechter}{Press \&
  Schechter}{1974}]{press:1974}
Press W.~H.,  Schechter P., 1974, ApJ, 187, 425

\bibitem[\protect\citeauthoryear{Primack et~al.}{Primack
  et~al.}{1995}]{primack:1995a}
Primack J.~R., Holtzman J., Klypin A.,  Caldwell D.~O., 1995, Phys.\ Rev.\
  Lett., 74, 2160

\bibitem[\protect\citeauthoryear{Quinn et~al.}{Quinn et~al.}{1997}]{quinn:1997}
Quinn T., Katz N., Stadel J.,  Lake G., 1997, preprint, astro-ph/9710043

\bibitem[\protect\citeauthoryear{Reid}{Reid}{1997}]{reid:1997}
Reid I.~N., 1997, AJ, 114, 161

\bibitem[\protect\citeauthoryear{Scott et~al.}{Scott et~al.}{1996}]{scott:1996}
Scott P.~F. et~al., 1996, ApJ, 461, L1

\bibitem[\protect\citeauthoryear{Seljak \& Zaldarriaga}{Seljak \&
  Zaldarriaga}{1996}]{seljak:1996}
Seljak U.,  Zaldarriaga M., 1996, ApJ, 469, 437

\bibitem[\protect\citeauthoryear{Smith et~al.}{Smith et~al.}{1998}]{smith:1998}
Smith C.~C., Klypin A., Gross M.~A.~K., Primack J.~R.,  Holtzman J., 1998,
  MNRAS, accepted, astro-ph/9702099

\bibitem[\protect\citeauthoryear{Smoot et~al.}{Smoot et~al.}{1992}]{smoot:1992}
Smoot G.~F. et~al., 1992, ApJ, 396, L1

\bibitem[\protect\citeauthoryear{Somerville \& Primack}{Somerville \&
  Primack}{1998}]{somerville:1998b}
Somerville R.,  Primack J., 1998, MNRAS, submitted, astro-ph/9802268, SP98

\bibitem[\protect\citeauthoryear{Somerville}{Somerville}{1997}]{somerville:the%
sis}
Somerville R.~S., 1997, Ph.D. thesis, University of California, Santa Cruz,
  http://www.fiz.huji.ac.il/~rachels/thesis.html

\bibitem[\protect\citeauthoryear{Somerville et~al.}{Somerville
  et~al.}{1998}]{somerville:1998a}
Somerville R.~S., Lemson G., Kolatt T.~S.,  Dekel A., 1998, in preparation

\bibitem[\protect\citeauthoryear{Squires et~al.}{Squires
  et~al.}{1996}]{squires:1996}
Squires G., Kaiser N., Babul A., Fahlman G., Woods D., Neumann D.~M.,
  B\"oringer H., 1996, ApJ, 461, 572

\bibitem[\protect\citeauthoryear{Squires et~al.}{Squires
  et~al.}{1997}]{squires:1997}
Squires G., Neumann D.~M., Kaiser N., Arnaud M., Babul A., B\"ohringer H.,
  Fahlman G.,  Woods D., 1997, apj, 482, 482

\bibitem[\protect\citeauthoryear{Steidel et~al.}{Steidel
  et~al.}{1998}]{steidel:1998}
Steidel C.~C., Adelberger K.~L., Dickinson M., Giavalisco M., Pettini M.,
  Kellogg M., 1998, ApJ, 492, 428

\bibitem[\protect\citeauthoryear{Summers, Davis, \& Evrard}{Summers
  et~al.}{1995}]{summers:1995}
Summers F.~J., Davis M.,  Evrard A.~E., 1995, ApJ, 454, 1

\bibitem[\protect\citeauthoryear{Tegmark}{Tegmark}{1996}]{tegmark:1996}
Tegmark M., 1996, ApJ, 464, L35

\bibitem[\protect\citeauthoryear{Tytler, Fan, \& Burles}{Tytler
  et~al.}{1996}]{tytler:1996}
Tytler D., Fan X.,  Burles S., 1996, Nat, 381, 207

\bibitem[\protect\citeauthoryear{Viana \& Liddle}{Viana \&
  Liddle}{1996}]{viana:1996}
Viana P.~T.~P.,  Liddle A.~R., 1996, MNRAS, 281, 323

\bibitem[\protect\citeauthoryear{Walter \& Klypin}{Walter \&
  Klypin}{1996}]{walter:1996}
Walter C.,  Klypin A., 1996, ApJ, 462, 13

\bibitem[\protect\citeauthoryear{Wechsler et~al.}{Wechsler
  et~al.}{1998}]{wechsler:1998}
Wechsler R.~H., Gross M.~A.~K., Primack J.~R., Blumenthal G.~R.,  Dekel A.,
  1998, ApJ, in press, astro-ph/9712141

\bibitem[\protect\citeauthoryear{White, Efstathiou, \& Frenk}{White
  et~al.}{1993}]{white:1993}
White S.~D.~M., Efstathiou G.,  Frenk C.~S., 1993, MNRAS, 262, 1023 (WEF93)

\bibitem[\protect\citeauthoryear{Wu \& Fang}{Wu \& Fang}{1996}]{wu:1996}
Wu~X.~P.,  Fang L.~Z., 1996, ApJ, 467, L45

\bibitem[\protect\citeauthoryear{Wu \& Fang}{Wu \& Fang}{1997}]{wu:1997}
Wu~X.~P.,  Fang L.~Z., 1997, ApJ, 483, 62

\bibitem[\protect\citeauthoryear{Xu}{Xu}{1995}]{xu:1995}
Xu~G., 1995, ApJS, 98, 355

\bibitem[\protect\citeauthoryear{Yepes et~al.}{Yepes et~al.}{1997}]{yepes:1997}
Yepes G., Kates R., Khokhlov A.,  Klypin A., 1997, MNRAS, 284, 235

\bibitem[\protect\citeauthoryear{Zel'dovich}{Zel'dovich}{1970}]{zeldovich:1970}
Zel'dovich Y.~B., 1970, A\&A, 5, 84

\end{thebibliography}
\fi




\ifref
\pagestyle{empty}
\thispagestyle{empty}

\renewcommand{\thefigure}{\arabic{figure}}
\renewcommand{\thetable}{\arabic{table}}

\processdelayedfloats
\let\small=\Large
\setcounter{figure}{0}
\begin{figure*}
  \Large
  \begin{figwide}{ps}
  \end{figwide}
\end{figure*}

\processdelayedfloats
\begin{figure*}
  \Large
  \begin{figwide}{cmbcmp}
  \end{figwide}
\end{figure*}

\processdelayedfloats
\begin{figure*}
  \Large
  \begin{figwide}{power_lin}
  \end{figwide}
\end{figure*}

\processdelayedfloats
\begin{figure*}
  \Large
  \begin{figwide}{power_bf}
  \end{figwide}
\end{figure*}

\processdelayedfloats
\begin{figure*}
  \Large
  {\Large\resizebox*{\linewidth}{!}{{
\setlength{\unitlength}{0.1bp}
\begin{picture}(4320,1512)(0,0)
\special{psfile=bias.eps llx=0 lly=0 urx=864 ury=353 rwi=8640}
\put(2385,1412){\makebox(0,0){Nonlinear power spectra}}
\put(100,631){%
\special{ps: gsave currentpoint currentpoint translate
270 rotate neg exch neg exch translate}%
\makebox(0,0)[b]{\shortstack{$b=\sqrt{\frac{P_{\mathrm{APM}}(k)}{P(k)}}$}}%
\special{ps: currentpoint grestore moveto}%
}
\put(450,1136){\makebox(0,0)[r]{1.4}}
\put(450,883){\makebox(0,0)[r]{1.2}}
\put(450,631){\makebox(0,0)[r]{1}}
\put(450,379){\makebox(0,0)[r]{0.8}}
\put(450,126){\makebox(0,0)[r]{0.6}}
\end{picture} }}}
  \begin{figwide}{power_nl}
  \end{figwide}
\end{figure*}

\processdelayedfloats
\begin{figure*}
  \Large
  \begin{figwide}{power_bw}
  \end{figwide}
\end{figure*}

\processdelayedfloats
\begin{figure*}
  \Large
  \begin{figwide}{power_rs}
  \end{figwide}
\end{figure*}

\processdelayedfloats
\begin{figure*}
  \Large
  \begin{figwide}{restest}
  \end{figwide}
\end{figure*}

\processdelayedfloats
\begin{figure*}
  \Large
  \begin{figwide}{mftest}
  \end{figwide}
\end{figure*}

\processdelayedfloats
\begin{figure*}
  \Large
  \begin{figlong}{mf}
  \end{figlong}
\end{figure*}

\processdelayedfloats
\begin{figure*}
  \Large
  \begin{figwide}{delc}
  \end{figwide}
\end{figure*}

\processdelayedfloats
\begin{figure*}
  \Large
  \begin{figwide}{cf}
  \end{figwide}
\end{figure*}

\fi
\end{document}